\documentclass[sigconf]{acmart}

\AtBeginDocument{%
  \providecommand\BibTeX{{%
    \normalfont B\kern-0.5em{\scshape i\kern-0.25em b}\kern-0.8em\TeX}}}

\setcopyright{acmcopyright}
\copyrightyear{2023}
\acmYear{2023}
\acmDOI{XXXXXXX.XXXXXXX}

\acmConference[SIGMOD '23]{ACM SIGMOD International Conference on Management of Data}{June 18--23,
  2023}{Seattle, WA}
\acmPrice{15.00}
\acmISBN{978-1-4503-XXXX-X/18/06}

\usepackage[ruled,linesnumbered,noend]{algorithm2e}
\usepackage{enumitem}
\usepackage{dsfont}
\usepackage{dutchcal}
\usepackage{cancel}
\usepackage[normalem]{ulem}
\usepackage{makecell}
\usepackage{multirow}

\newcommand{\oursolution}{FIRM}
\begin{document}
\setlength\abovedisplayskip{-0.3pt}
\setlength\belowdisplayskip{-0.3pt}
\setlength\abovedisplayshortskip{-0.3pt}
\setlength\belowdisplayshortskip{-0.3pt}
\setlength{\textfloatsep}{2pt}

\setlength{\floatsep}{0.5pt}

\setlength{\abovecaptionskip}{0.5pt}
\setlength{\belowcaptionskip}{0.5pt}
\setlength{\dbltextfloatsep}{0.5pt}
\title{Personalized PageRank on Evolving Graphs with an Incremental Index-Update Scheme}

\author{Guanhao Hou}
\email{ghhou@se.cuhk.edu.hk}
\affiliation{%
  \institution{The Chinese University of Hong Kong}
  \country{Hong Kong SAR}
}

\author{Qintian Guo}
\email{qtguo@se.cuhk.edu.hk}
\affiliation{%
  \institution{The Chinese University of Hong Kong}
  \country{Hong Kong SAR}
}

\author{Fangyuan Zhang}
\email{fzhang@se.cuhk.edu.hk}
\affiliation{%
  \institution{The Chinese University of Hong Kong}
  \country{Hong Kong SAR}
}

\author{Sibo Wang}
\email{swang@se.cuhk.edu.hk}
\affiliation{%
  \institution{The Chinese University of Hong Kong}
  \country{Hong Kong SAR}
}

\author{Zhewei Wei}
\email{zhewei@ruc.edu.cn}
\affiliation{%
  \institution{Renmin University of China}
  \city{Beijing}
  \country{China}
}

\renewcommand{\shortauthors}{Hou, et al.}

\begin{abstract}
{\em Personalized PageRank (PPR)} stands as a fundamental proximity measure in graph mining. Given an input graph $G$ with the probability of decay $\alpha$, a source node $s$ and a target node $t$, the PPR score $\pi(s,t)$ of target $t$ with respect to source $s$ is the probability that an $\alpha$-decay random walk starting from $s$ stops at $t$. A {\em single-source PPR (SSPPR)} query takes an input graph $G$ with decay probability $\alpha$ and a source $s$, and then returns the PPR $\pi(s,v)$ for each node $v \in V$. Since computing an exact SSPPR query answer is prohibitive, most existing solutions turn to approximate queries with guarantees. The state-of-the-art solutions for approximate SSPPR queries are index-based and mainly focus on static graphs, while real-world graphs are usually dynamically changing. However, existing index-update schemes can not achieve a sub-linear update time.

Motivated by this, we present an efficient indexing scheme for single-source PPR queries on evolving graphs. Our proposed solution is based on a classic framework that combines the forward-push technique with a random walk index for approximate PPR queries. Thus, our indexing scheme is similar to existing solutions in the sense that we store pre-sampled random walks for efficient query processing. One of our main contributions is an incremental updating scheme to maintain indexed random walks in expected $O(1)$ time after each graph update. To achieve $O(1)$ update cost, we need to maintain auxiliary data structures for both vertices and edges. To reduce the space consumption, we further revisit the sampling methods and propose a new sampling scheme to remove the auxiliary data structure for vertices while still supporting $O(1)$ index update cost on evolving graphs. Extensive experiments show that our update scheme achieves orders of magnitude speed-up on update performance over existing index-based dynamic schemes without sacrificing the query efficiency.

\end{abstract}

\begin{teaserfigure}
\end{teaserfigure}

\maketitle

\def\header{\vspace{1mm} \noindent}
\def\done{\hspace*{\fill} $\framebox[2mm]{}$ \medskip}

\def\diffsym{{{}_\Delta}}
\newcommand{\tuple}[1]{\left\langle #1 \right\rangle}
\newcommand{\vect}[1]{\boldsymbol{#1}}
\newcommand{\mceil}[1]{\left\lceil #1 \right\rceil}
\newcommand{\mfloor}[1]{\left\lfloor #1 \right\rfloor}

\newcommand{\bigO}[1]{O\left(#1\right)}
\newcommand{\bigTheta}[1]{\Theta\left(#1\right)}
\newcommand{\bigOmega}[1]{\Omega\left(#1\right)}

\def\E{\mathbb{E}}
\def\Pr{\mathbb{P}}
\def\pspace{\mathcal{S}}
\def\evspace{\mathcal{F}}
\newcommand{\1}[1]{\mathds{1}_{\left[#1\right]}}

\def\Ngh{\mathcal{N}}
\def\Edge{\mathcal{E}}

\def\taume{{\tau\text{-}1}}
\newcommand{\past}[1]{#1_\taume}
\newcommand{\cur}[1]{#1_\tau}

\def\emalpha{{1\mathrm{-}\alpha}}
\def\rmax{r_{max}}
\def\pih{\hat{\pi}}
\def\pitil{\tilde{\pi}}
\newcommand{\funcname}[1]{\operatorname{\bf #1}}

\newenvironment{proofsketch}{%
  \renewcommand{\proofname}{Proof Sketch}\proof}{\endproof}

\def\agendastar{$\textrm{Agenda}^\#$}

\newcommand{\trver}[1]{{#1}}
\newcommand{\crver}[1]{}

\section{Introduction}
Given an input graph $G$, a decay probability $\alpha$, a source node $s$, and a target node $t$, the {\em personalized PageRank (PPR)} score $\pi(s,t)$ stands as a classic proximity measure of the relative importance of $t$ from the viewpoint of $s$. More formally, the PPR score $\pi(s,t)$ of $t$ with respect to $s$ is the probability that an $\alpha$-decay random walk~\cite{JehW03} starting from $s$ stops at $t$. Here, an {\em $\alpha$-decay random walk} works as follows: it starts from the source $s$ and at each step, it either {\em (i)} stops at the current node $v$ (initially $v=s$) with probability $\alpha$; or {\em (ii)} randomly jumps to one of the out-neighbors of $v$ with probability $\emalpha$ (specially, a self-loop will be applied to $v$ if it has no out-neighbor). An important type of PPR queries is the {\em single-source PPR (SSPPR)} query where a source $s$ is given, and the goal is to compute the PPR score $\pi(s, v)$ for each node $v$ in the input graph $G$. SSPPR has many important applications in web search~\cite{Page1999,JehW03}, spam detection~\cite{GBGP06}, community detection~\cite{AnCL06}, social recommendations~\cite{GGLSWZ13, Lin19}, etc. Moreover, as a fundamental proximity measure, SSPPR is also widely used in graph representation learning, e.g., HOPE~\cite{OuCPZZ16}, Verse~\cite{TsiMKM18}, STRAP~\cite{YinWei19}, NRP \cite{YangSXYB20}, PPRGo~\cite{BojKPKBR20}, and DynamicPPE~\cite{GuoZS21}.

With its wide applications, it is important to have efficient algorithms for SSPPR queries. However, computing exact answers for SSPPR queries is prohibitive on massive graphs. Existing state-of-the-art solutions~\cite{LinWXW20,WuGWZ21} all focus on {\em approximate SSPPR (ASSPPR)} queries with guarantees (Ref. to Definition \ref{def::assppr}). With the framework which combines the Forward-Push technique with the Monte-Carlo method for random walks, existing solutions, e.g., FORA+~\cite{WYWXWLYT19} and SpeedPPR+~\cite{WuGWZ21}, pre-store random walks and show superb query efficiency on ASSPPR queries while providing strong approximation guarantees. However, all these index-based methods assume that the input graph is static while most real-world graphs are dynamically evolving. Social recommendation systems, e.g. Twitter Whom-to-Follow~\cite{GGLSWZ13} and Tencent social recommendation~\cite{Lin19} are virtually required to work on social networks which keep evolving frequently. Besides, some graph representation learning frameworks, e.g. SDG~\cite{FuH21}, need to estimate SSPPR on evolving graphs. Despite that index-free solutions can also handle ASSPPR queries on dynamic graphs, existing solutions with random walk index have a significant improvement in query performance compared to index-free solutions. In addition, early index-based solutions for ASSPPR queries on dynamic graphs~\cite{BahCG10,ZhLG16} have a prohibitive cost to maintain index structures. Even the state-of-the-art index-update solution, Agenda~\cite{MoLuo21}, still has a time consumption that is linearly related to the graph size for each update.

Motivated by limitations of existing solutions, we propose an efficient index scheme {\oursolution}\footnote{\underline{F}orward-Push with \underline{I}ncremental \underline{R}andom Walk \underline{M}aintenance}, which solves single-source PPR queries on evolving graphs. Our proposed solution is also based on the classic Forward-Push + Monte-Carlo framework, and thus it can provide query processing as efficiently as existing solutions on static graphs. One of our main contributions is that we design an incremental index update scheme to maintain the random walk index after each graph update in expected $O(1)$ time. The main idea is to find affected random walks for the recent graph update and adjust such random walks to fit the updated graph. We design auxiliary structures to trace affected random walks efficiently and achieve the $O(1)$ expected update time. In particular, we maintain auxiliary data structures for each node and each edge which record the information of random walks that go across each node and each edge, respectively. To reduce the space consumption, we further revisit the sampling method and present a non-trivial new sampling scheme to remove the requirement of the auxiliary data structure for each node. Remarkably, the new scheme comes at no sacrifice of the update time complexity as it still provides $O(1)$ expected time cost for each graph update. 
Extensive experiments show that our update scheme achieves up to orders of magnitude speed-up over existing index-based schemes on evolving graphs without sacrificing the query efficiency over index-based solutions.

\section{Problem Definition}
Let $G = (V,E)$ be a directed graph with node number $n \mathrm{=} |V|$ and edge number $m \mathrm{=} |E|$. For an edge $e \mathrm{=} \tuple{u,v} \in E$, we say that $e$ is an outgoing edge of node $u$, and node $v$ is an out-neighbor of node $u$. Then for a node $u \in V$, the set $\Ngh(u)$ of its out-neighbors, the set $\Edge(u)$ of its outgoing edges, and the out-degree $d(u)$ of node $u$ are denoted as $\Ngh(u) = \{v | \tuple{u,v} \in E \}$, $\Edge(u) = \{ \tuple{u,v} \in E \}$, and $d(u) \mathrm{=} |\Edge(u)|$, respectively.

\header
{\bf Personalized PageRank (PPR)~\cite{JehW03}.} Given a directed graph $G = (V,E)$, a source node $s \in V$ and a decay probability $\alpha$, the PPR score of node $t$ with respect to node $s$ is defined as the probability that an $\alpha$-decay random walk starting from $s$ stops at $t$, denoted as $\pi(s,t)$. We use a vector $\vect{\pi}(s)$ to represent the PPR scores of all nodes in the graph with respect to source node $s$.

\header
{\bf Approximate Single-Source PPR (ASSPPR).} Following ~\cite{WYXWY17,WuGWZ21,MoLuo21}, this paper focuses on the ASSPPR problem defined as follows:

\vspace{-1.5mm}
\begin{definition} [$(\epsilon,\delta)$-ASSPPR] \label{def::assppr}
Given a threshold $\delta \mathrm{\in} (0,1]$, a relative error $\epsilon \mathrm{\in} (0,1)$, and a failure probability $p_f$, an ASSPPR query returns an estimated PPR score $\pitil(s, v)$ for each node $v \in V$ so that:
\begin{equation*}
  |\pi(s, v) - \pitil(s, v)| < \epsilon \cdot \pi(s, v) \quad \forall \pi(s, v) \ge \delta
\end{equation*}
with at least $1\mathrm{-}p_f$ probability where the parameters $\delta$ and $p_f$ are typically set to $O(1\mathrm{/}n)$.  \done
\end{definition}

\vspace{-1.5mm}
Besides, many applications, e.g. recommendation system \cite{GGLSWZ13, Lin19} and graph neural networks \cite{BojKPKBR20, FuH21}, are not interested in the full PPR vector with respect to $s$. Instead, they only use the vertices which have the highest PPR scores with respect to $s$. Thus, we study another type of PPR queries, called {\em approximate single-source top-$k$ PPR (ASSPPR top-$k$)} queries~\cite{WYXWY17} defined as follows:

\vspace{-1.5mm}
\begin{definition} \label{def:asspprtk}($(\epsilon,\delta)$-ASSPPR top-$k$) Given a threshold $\delta \mathrm{\in} (0,1]$, a relative error $\epsilon \mathrm{\in} (0,1)$, a failure probability $p_f$, and a positive integer $k$, an ASSPPR top-$k$ query returns a sequence of $k$ nodes, $v_1 , v_2, \dots, v_k$, such that for any $i \in [1..k]$, the following equations hold with at least $1\mathrm{-}p_f$ probability:
\begin{equation*}
\left.\begin{aligned}
&|\pi(s,v_i) - \pitil(s,v_i)| \le \epsilon \cdot \pi(s,v_i) \\
&\pi(s,v_i) \ge (1-\epsilon) \cdot \pi(s,v_i^*)
\end{aligned} \right.
\quad \forall \pi(s,v_i^*) \ge \delta,
\end{equation*}
where $v_i^*$ has the $i$-th largest exact PPR score with respect to $s$.  \done
\end{definition}

\begin{table}[!t]
\centering
\begin{small}
\renewcommand{\arraystretch}{1.2}
\caption{Frequently used notations.} \label{tab:notations}
\begin{tabular}{|p{0.2\columnwidth}|p{0.72\columnwidth}|}
\hline
{\bf Notation} & {\bf Description} \\
\hline
$G=(V,E)$ & a directed graph consists of node set $V$ and edge set $E$ \\
\hline
$n,m$ & the number of nodes and edges, respectively \\
\hline
$\Ngh(u)$ & the set of out-neighbors of node $u$ \\
\hline
$\Edge(u)$ & the set of outgoing edges of node $u$ \\
\hline
$d(u)$ & the out-degree of node $u$ \\
\hline 
$\cur{G}$$=$$(\cur{V},\cur{E})$ & the graph after the $\tau$-th update has been executed \\
\hline
$\cur{e}$$=$$\tuple{\cur{u},\cur{v}}$ & the $\tau$-th updating edge, which occurs at timestamp $\tau$ \\
\hline
$\pi(s,v)$ & the PPR score for node $v$ with respect to source node $s$ \\
\hline
$H$, $H(s)$ & the pre-stored random walk index and those starting from node $s$, respectively \\
\hline
$\pitil(s,v)$ & the estimation of the PPR score $\pi(s,v)$ \\
\hline
$\pih(s,v),r(s,v)$ & the reserve and residue value of node $v$ with respect to source node $s$, respectively \\
\hline
$\epsilon$, $\delta$, $p_f$ & relative error bound, effective threshold, and failure probability, respectively \\
\hline
\end{tabular}
\end{small}
\vspace{0mm}
\end{table}
\noindent
{\bf Evolving Graph.} We make a consistent assumption of the dynamically evolving graph with previous works \cite{BahCG10, ZhLG16,MoLuo21}. There exist an initial graph at the beginning, followed by a sequence of updates. In this paper, we only consider edge updates: \underline{\em edge insertion} and \underline{\em edge deletion}. There may also exist node insertions/deletions on evolving graphs, both of which can be easily converted to a sequence of edge insertions/deletions. We add a subscript to specify which timestamp we are discussing. Let $G_0$ be the initial graph, $\cur{e}$ denotes the $\tau$-th updating edge (the update is either an insertion or a deletion). Let $\cur{G}=(\cur{V},\cur{E})$ denotes the graph after the $\tau$-th update has been applied. Note that in the insertion case, we have $\cur{E} = \past{E} \mathrm{\cup} \{ \cur{e} \}$; in the deletion case, we have $\cur{E} = \past{E} \mathrm{\setminus} \{ \cur{e} \}$. Besides, we consider that the $\tau$-th update occurs at timestamp $\tau$, and thus the update at timestamp $\tau$ refers to the $\tau$-th update.

With the edge updating model, we further assume that the sequence of updates is uniformly random, which is called \textit{random arrival model} in \cite{BahCG10, ZhLG16}. More precisely, it is defined as follows.

\vspace{-1.5mm}
\begin{definition}[Random arrival model]\label{def:ram}
For edge insertion, the probability that $e \mathrm{=} \tuple{u,v}$ is inserted at timestamp $\tau$ is: 
\begin{equation*}
      \Pr\left[ \cur{e} = e \right] = \frac{1}{\cur{m}},
\end{equation*}
i.e., the probability that each edge in $\cur{E}$ is the last one to be inserted at timestamp $\tau$ is equal.
In the case of edge deletion, the probability that any edge $e \mathrm{=} \tuple{u,v} \in \past{E}$ will be deleted at timestamp $\tau$ is
\begin{equation*}
  \Pr\left[ \cur{e} = e \right] = \frac{1}{\past{m}},
\end{equation*} 
i.e., each edge has equal chance to be deleted at timestamp $\tau$. 
\done
\end{definition}

\vspace{-1.5mm}
From the above definition, we could immediately know that the probability that the updating edge $\cur{e}$ is an outgoing edge of node $u$ in the edge insertion case is:
\begin{equation}
  \Pr\left[ \cur{e} \in \cur{\Edge}(u) \right] = \frac{\cur{d}(u)}{\cur{m}}, \label{eqn:pr:ins} 
\end{equation}
and for the edge deletion case, the corresponding probability is
\begin{equation}
  \Pr\left[ \cur{e} \in \past{\Edge}(u) \right] = \frac{\past{d}(u)}{\past{m}}. \label{eqn:pr:del} 
\end{equation}

In the rest of this paper, for the sake of brevity, we use the non-subscripted notations (e.g., $G$, $E$, $d$) in the context if there is no danger of confusion. Otherwise, the subscripted notations are used. In Table \ref{tab:notations}, we list the notations frequently used in this paper.

\section{Existing Solutions}

In this section, we first revisit two index-free solutions for ASSPPR queries on static graphs: FORA~\cite{WYWXWLYT19} and SpeedPPR~\cite{WuGWZ21} and their index-based solution FORA+ and SpeedPPR+. Then, we review Agenda~\cite{MoLuo21}, the state-of-the-art solution for evolving graphs.

\subsection{Solutions on Static Graph} \label{sec:sol:static}

{\bf FORA.} Wang et al.~\cite{WYWXWLYT19} propose a two-phase solution, dubbed as FORA, to answer ASSPPR queries. It first performs the Forward-Push technique~\cite{AnCL06} as Algorithm~\ref{algo:push} shows.
In particular, it maintains two vectors, the reserve vector $\boldsymbol{\pih(s)}$ and the residue vector $\boldsymbol{r(s)}$. Initially, the vector $\boldsymbol{\pih(s)}$ is set to zero on all entries and we use $\pih(s,v)$ to denote the $v$-th entry of vector $\boldsymbol{\pih(s)}$. For the residue vector $\boldsymbol{r(s)}$, it is initialized as $\vect{1}_s$ where $\vect{1}_s$ denotes the one-hot vector with respect to $s$, i.e., only $r(s,s)$ is 1 and other positions are zero. Then for any node $u$ with residue no smaller than $\rmax \mathrm{\cdot} d(u)$, it performs a {\em push operation} to $u$, which converts $\alpha$ portion of the current residue to the reserve of $u$ (Line 3), and for remaining residues, they are evenly propagated to out-neighbors of $u$ (Lines 4-5). It stops when no node satisfies the push condition (Line 2). After the Forward-Push phase, it simulates sufficient random walks to give the final result. The rationale of FORA comes from the following invariant which always holds during the Forward-Push phase:
\begin{equation} \label{eqn:fora:fpinvar}
    \pi(s,t) = \pih(s,t) + \sum_{v \in V} r(s,v) \cdot \pi(v,t) 
\end{equation}
Therefore, FORA computes $\pih(s,t)$ to roughly approximate the PPR score in the first phase, and then exploits random walks to estimate the cumulative term $\sum_{v \in V} r(s,v) \mathrm{\cdot} \pi(v,t)$ in the second phase, thus refining the estimation. The following lemma about the number of random walks is proved in \cite{WYWXWLYT19}.

\vspace{-1mm}
\begin{lemma} \label{lm:fora:rwnum}
Given the threshold of residue $\rmax$, the number of independent random walks starting from node $v \in V$ should be at least $\mceil{r(s,v) \mathrm{\cdot} \omega}$ to satisfy $(\epsilon, \delta)$-approximate guarantee, where 
\begin{equation} \label{eqn:fora:omega}
\omega = \frac{\left((2/3)\cdot\epsilon+2\right) \cdot \log{(2/p_f)}}{\epsilon^2\delta}.
\end{equation}
\end{lemma}

\begin{algorithm}[t]
\caption{Forward-Push}\label{algo:push}
\LinesNumbered
\KwIn{Graph $G=(V,E)$, decaying rate $\alpha$, threshold $\rmax$, source node $s$}
\KwOut{reserve vector $\vect{\pih}(s)$ and residue vector $\vect{r}(s)$}
$\vect{\pih}(s) \leftarrow \vect{0}, \vect{r}(s) \leftarrow \vect{1}_s$\;
\While {$\exists u \in V$ {\bf such that} $\frac{r(s,u)}{d(u)} \geq \rmax$ }{
  $\pih(s,u) \leftarrow \pih(s,u) + \alpha \cdot r(s,u)$\;
  \ForEach{$v \in \Ngh(u)$}{
    $r(s,v) \leftarrow r(s,v) + (\emalpha) \cdot \frac{r(s,u)}{d(u)}$\;
  }
  $r(s,u) \leftarrow 0$\;
}
\Return{$[\vect{\pih}(s), \vect{r}(s)]$}\;
\end{algorithm}

As \cite{AnCL06} shows that the time complexity of Forward-Push phase corresponding to the threshold of residue $\rmax$ is $O(1\mathrm{/}\rmax)$, FORA sets $\rmax \mathrm{=} \sqrt{1\mathrm{/}(m\mathrm{\cdot}\omega)}$ to optimize the time complexity to $O(\sqrt{m\mathrm{\cdot}\omega})$. Since a scale-free graph with $\gamma \mathrm{\in} [2, 3]$ has the average degree $\bar{d} \mathrm{=} m\mathrm{/}n \mathrm{=} O(\log{n})$, the above time complexity will be $O(n\mathrm{\cdot}\log{n}\mathrm{/}\epsilon)$ with $\delta \mathrm{=} 1\mathrm{/}n$ and $p_f \mathrm{=} 1\mathrm{/}n$. Interested readers can refer to~\cite{WYWXWLYT19} for more details of the time complexity.

\header
{\bf FORA+.} Wang et al.~\cite{WYWXWLYT19} also propose an index scheme of FORA that works on static graphs, dubbed as FORA+, which pre-computes a sufficient number of random walks to further improve the query efficiency of their algorithm. For each random walk, FORA+ simply stores the source and terminal node. Then, when FORA+ needs to sample a random walk starting from a node $u$, it directly gets a pre-stored random walk with $u$ as the source and chooses a pre-stored terminal node. Notice that each pre-stored random walk will be used at most once for an ASSPPR query to guarantee that random walks are independent from each other. One important question for FORA+ is how many random walks should be pre-computed for each node $v \in V$ such that it could achieve both accuracy guarantee and space efficiency. According to Lemma \ref{lm:fora:rwnum}, they have:
\begin{lemma} \label{lm:fora:index}
Given the threshold of residue $\rmax$, the number of pre-computed independent random walks starting from node $v \in V$ should be at least $\mceil{d(v)\mathrm{\cdot}\rmax\mathrm{\cdot}\omega}$ to satisfy $(\epsilon,\delta)$-approximate guarantee.
\end{lemma}
By setting $\rmax \mathrm{=} \sqrt{1\mathrm{/}(m\mathrm{\cdot}\omega)}$ as in FORA, FORA+~\cite{WYWXWLYT19} has a space complexity of $\Theta(m\mathrm{+}\sqrt{m\mathrm{\cdot}\omega})$ and will become $O(n\mathrm{\cdot}\log{n}\mathrm{/}\epsilon)$ on scale-free graphs with $\gamma \mathrm{\in} [2,3]$.

\header
{\bf SpeedPPR.} Wu et al.~\cite{WuGWZ21} propose an improved version of FORA, named SpeedPPR, to answer ASSPPR queries. In practice, SpeedPPR takes benefits from their cache-friendly implementation of Forward-Push, called Power-Push, which combines the power iteration and vanilla Forward-Push into a whole. As well as they theoretically prove that the Forward-Push phase could achieve a time complexity of $O(m\mathrm{\cdot}\log{(1\mathrm{/}\rmax)})$ apart from the previously known result $O(1\mathrm{/}\rmax)$ and this bound improves when $\rmax$ is very small, i.e., $O(1\mathrm{/}m)$. With the aid of such new knowledge, they further set the threshold $\rmax \mathrm{=} \Theta(1\mathrm{/}\omega)$ in SpeedPPR, rather than $\sqrt{1\mathrm{/}(m\mathrm{\cdot}\omega)}$ in the original FORA. It is proved in~\cite{WuGWZ21} that such a modification yields a time complexity of $O(n\mathrm{\cdot}\log{n}\mathrm{\cdot}\log{(1\mathrm{/}\epsilon)})$ on scale-free graphs, improving a factor of $1\mathrm{/}(\epsilon\mathrm{\cdot}\log{(1\mathrm{/}\epsilon)})$ over FORA. Interested readers can refer to~\cite{WuGWZ21} for more details of the time complexity.

\header
{\bf SpeedPPR+}. SpeedPPR also admits an index-based version, called SpeedPPR+~\cite{WuGWZ21}. It is showed in~\cite{WuGWZ21} that as $\rmax\mathrm{\cdot}\omega \mathrm{=} \Theta(1)$, the index size of SpeedPPR+ is $\Theta(m)$, independent of the relative error bound $\epsilon$ and the effective threshold $\delta$, as well small than the space complexity of FORA+ which is $\Theta(m\mathrm{+}\sqrt{m\mathrm{\cdot}\omega})$.

\subsection{Solutions on Evolving Graph} \label{sec:sol:dyn}
There is no doubt that pure-online approaches such as FORA can be applied to dynamically evolving graphs directly. Nevertheless, the previous works~\cite{WYWXWLYT19,WuGWZ21} show that an index-based scheme is much more efficient in query processing than its index-free version. On the other hand, there is also a trivial tactic to adapt indexing schemes to evolving graphs which reconstructs the whole index after every update. However, such a strategy is obviously inefficient in dynamic graphs while most real-world graphs evolve frequently.

The early index-based solutions for ASSPPR queries on evolving graphs try to maintain random walks~\cite{BahCG10} or the reserve and residue vectors~\cite{ZhLG16} incrementally, while the index cost and update cost are still inconceivable to answer ASSPPR queries for arbitrary source nodes. For instance, the solution in \cite{BahCG10} needs $O(n^2)$ space to pre-store random walks to answer ASSPPR queries for arbitrary source nodes when $\delta\mathrm{=}1\mathrm{/}n$, making them infeasible to large graphs.

\header
{\bf Agenda.} Mo and Luo~\cite{MoLuo21} propose a feasible approach to maintain the index of FORA on evolving graphs named Agenda, which aims to balance the query and update efficiencies of ASSPPR estimation. The core strategy of Agenda is lazy-update, which has a tolerance for the inaccuracy of random walks and reconstructs a part of random walks when the error exceeds the limit. 

Agenda introduces a parameter $\theta \mathrm{\in} (0,1)$ and makes an inaccuracy tolerance of its index. When an edge update $\cur{e}\mathrm{=}\tuple{\cur{u},\cur{v}}$ comes, it traces the (upper bound of) inaccuracy of current index by performing a Backward-Push~\cite{AnBCHMT07} starting from $\cur{u}$, and accumulates the inaccuracy of each node into a vector $\vect{\sigma}$. For query processing, Agenda splits the error tolerance. Specifically, it first invokes the Forward-Push phase of FORA whereas the $\rmax$ is set according to the relative error bound $\theta\mathrm{\cdot}\epsilon$ instead of $\epsilon$. With this tightened error bound $\theta \mathrm{\cdot} \epsilon$, if there is no update at all, the query accuracy of Agenda tends to be higher than that of the original FORA/SpeedPPR. Then, it checks the query-dependent inaccuracy of current index with $\vect{e}\mathrm{=}\vect{\sigma} \mathrm{\circ} \vect{r}$ (where $\vect{r}$ is the residue vector and $\mathrm{\circ}$ is the pairwise multiplication). Next, it repeats to reconstruct all random walks starting from node $v$ which has the largest value in $\vect{e}$, until the inaccuracy of the index with respect to the query will not exceed a relative error $(1\mathrm{-}\theta)\mathrm{\cdot}\epsilon$. Finally, Agenda enters the refining phase of FORA with the less inaccurate index. Since the relative error of the two parts (i.e. the FORA process and the index itself) is bounded by  $\theta\mathrm{\cdot}\epsilon$ and $(1\mathrm{-}\theta)\mathrm{\cdot}\epsilon$, respectively, the final result can be bounded by relative error $\epsilon$, so that Agenda can answer an ASSPPR query with $(\epsilon,\delta)$-approximate guarantee.

It is proved in~\cite{MoLuo21} that the expected fraction of index to be reconstructed for each update will be either $O(\epsilon\mathrm{\cdot}\cur{\bar{d}}\mathrm{/}\log{\cur{n}})$ by setting $\rmax^b\mathrm{=}\Theta(1\mathrm{/}\cur{n})$ on directed graphs (where $\cur{\bar{d}}$ is the average degree of $\cur{G}$), or $O(\epsilon\mathrm{/}\log{\cur{n}})$ by setting $\rmax^b\mathrm{=}\Theta(\cur{d}(\cur{u})\mathrm{/}\cur{m})$ on undirected graphs. Agenda significantly reduces the number of random walks to be re-sampled for each update, whereas according to~\cite{AnBCHMT07}, backward-push expectantly runs within $O(\bar{d}\mathrm{/}\rmax^b)$ time for random starting node, and thus the time cost for tracing the inaccuracy is $O(\cur{m})$ for each update, which will also become more and more expensive when the graph is becoming larger and larger.

\header{\bf \agendastar.} Since Agenda divides the error tolerance into two parts, the error bound of its FORA process becomes $\theta\mathrm{\cdot}\epsilon$ which is tighter than that of the original FORA/SpeedPPR. That means Agenda requires a more precise result of Forward-Push phase, and thus it makes a trade-off between query performance and update cost. On the other hand, Agenda makes a conservative estimation to bound the error resulting from the remaining non-updated part of the index, which is $(1\mathrm{-}\theta)\mathrm{\cdot}\epsilon$. In practice, the actual inaccuracy of its index after the lazy-update process often becomes far smaller than the upper bound it reckons. Therefore, in most cases, Agenda wastes computation resources to provide over-exquisite results. In our experiments, we present a new version of Agenda, dubbed as {\agendastar}, which aggressively assumes that the inaccuracy (with respect to the processing query) after the lazy-update phase is negligible. Thus, We set the relative error of the FORA process to be $\epsilon$ instead of $\theta \mathrm{\cdot} \epsilon$. By such a strategy, the worst case relative error of {\agendastar} becomes $(2\mathrm{-}\theta)\mathrm{\cdot}\epsilon$. As we will see, {\agendastar} provides an accuracy pretty close to that of FORA/FORA+ when FORA/FORA+ provides a worst-case $\epsilon$-relative error. 

We include {\agendastar} as our baseline since we aim to provide a fair comparison where we compare the efficiency when all methods are providing a similar accuracy.

\section{The Incremental Approach} \label{sec:iprea}
Real-world graphs are usually dynamically evolving, which motivates us to find an efficient algorithm for ASSPPR problems on evolving graphs. Unfortunately, we need to make a great effort to achieve this goal. On one hand, the index-free solutions for static graphs, i.e., SpeedPPR, could easily handle the dynamic scenarios on their own right, but they may suffer from query efficiency issues since they need to perform a number of random walks for each query. On the other hand, as aforementioned,  existing index-based approaches for ASSPPR problems on evolving graphs, have a notable time cost to maintain their index. Even the state-of-the-art solution, Agenda, needs a time linearly correlating to the graph size for each update. Since many real-world graphs, especially the social networks are colossal, there will be a prohibitively computational overhead for index maintenance if update events happen frequently (unluckily, it is the usual case). Therefore, it deserves our effort to design a better index-based solution to tackle the challenges. 

\begin{algorithm}[t]
\caption{Update-Insert}\label{algo:ins}
\LinesNumbered
\KwIn{Graph $\cur{G} = (\cur{V}, \cur{E})$, past index $\past{H}$, edge $\cur{e} = \tuple{\cur{u},\cur{v}}$.}
\KwOut{New index $\cur{H}$ and $\cur{C^V}$.}
  $\cur{H} \leftarrow \past{H}$\;
  $C \leftarrow \funcname{Sample}(\past{C^V}(\cur{u}),  \frac{1}{\cur{d}(\cur{u})})$\; \label{algo:ins:upd}
  \ForEach {$c \in C$} {
    $w \leftarrow \cur{H}[c.id]$\;
      $w[c.step\mathrm{+}1] \leftarrow \cur{v}$\;\label{algo:ins:upd-redirected}
      $w \leftarrow \funcname{Walk-Restart}(\cur{G},w,c.step\mathrm{+}1)$\;\label{algo:ins:upd-continue}
  }\label{algo:ins:upd-end} 
  \While {$|\cur{H}(\cur{u})| <\mceil{\cur{d}(\cur{u}) \cdot \rmax \cdot \omega}$} { \label{algo:ins:add}
    $\cur{H} \leftarrow \cur{H} \cup \{\funcname{Random-Walk}(\cur{G},\cur{u})\}$\;
  }\label{algo:ins:add-end}
  \Return{$\cur{H}$}\;
\end{algorithm}

\header
{\bf Solution overview.} Generally speaking, our goal is also to find an efficient way to maintain the index of FORA+. However, different from Agenda, our proposal uses an eager-update strategy. Compared to the lazy-update strategy used by Agenda, our solution has a query efficiency almost the same as the solutions on static graphs because we make no trade-off between update and query efficiency. More importantly, by carefully tracing and minimally adjusting the random walks affected by each update, our solution has a constant time cost per update in expectation so that it can be applied to massive-scale and highly-frequently evolving graphs. 

To achieve efficient tracing, we should maintain more information on each random walk instead of its starting and terminal nodes. More specifically, each element $w$ of our index (denoted as $H$) is a completed path $\tuple{v_0,v_1,...,v_l}$ starting from $v_0$, crossing $v_1,...,v_{l\mathrm{-}1}$ in order, and terminating at $v_l$. Besides, we use $H(u)$ to denote the subset of $H$ that contains all random walks starting from $u$. Next, we will introduce details of our solutions, which take an expected constant index update time for both edge insertion and deletion. 

\header
{\bf Remark.} As a node without any incident edge on it has no effect on PPR values of other nodes, we can handle the node insertion case in this way: a new node is automatically inserted, exactly when the first edge which is incident on the new node arrives. Then, we proceed with remaining added edges following our edge insertion algorithm. In the case of node deletion, we can delete a node by deleting all edges incident to the node one by one, and the node will be deleted automatically once the last incident edge is removed.
 
\subsection{Edge Insertion}\label{sec:iprea:ins}

Intuitively, assuming that an edge $\tuple{\cur{u},\cur{v}}$ is inserted, we need to update the pre-stored random walks to guarantee that they are sampled according to the current graph structure. Meanwhile, the insertion only makes a slight change to the graph structure and hence it should affect only slightly the random walks. More specifically, we will show that if a pre-stored random walk does not cross node $\cur{u}$, then the random walk does not need to be updated. Notice that here we only consider random walks that cross $\cur{u}$ and do not include random walks that only terminate at $\cur{u}$ (but have not crossed $\cur{u}$). A quick explanation is that if the random walk just stops at node $\cur{u}$, which means it does not cross $\cur{u}$, it is still not affected as it is not dependent on the newly inserted edge $\tuple{\cur{u},\cur{v}}$.

\begin{figure}

\begin{small}
    \centering
    \begin{tabular}{p{0.33\columnwidth} p{0.33\columnwidth} p{0.33\columnwidth}}
        \includegraphics[height=20mm]{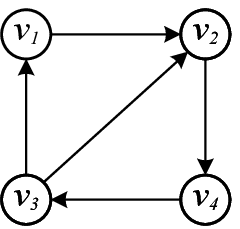} &
        \hspace{-3mm}\includegraphics[height=20mm]{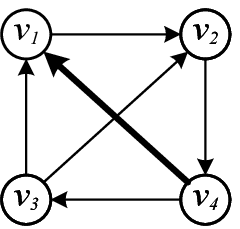} &
        \hspace{-3mm}\includegraphics[height=20mm]{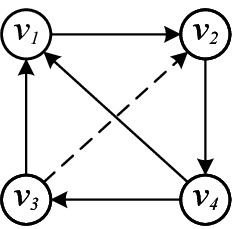}
        \\[-1mm]
       \hspace{0mm} (a) Initial graph $G_0$  &
        \hspace{-3mm} (b) $G_1$ after insertion & 
        \hspace{-3mm}(c) $G_2$ after deletion 
 \end{tabular}
\caption{A Tiny Evolving Graph.}\label{exm:graph}
\end{small}

\end{figure}
 
\begin{table}
\begin{small}

\centering
{
\renewcommand{\arraystretch}{1.1}
\begin{tabular}{|c|l|}
    \hline
    & ID: walking path \\
    \hline 
    $H_0(v_1)$ & $1: v_1 \rightarrow v_2 \rightarrow v_4,~2: v_1 \rightarrow v_2 $ \\
    \hline
    $H_0(v_2)$ & $3: v_2 \rightarrow v_4, ~4: v_2\rightarrow v_4 \rightarrow v_3 \rightarrow v_1$ \\
    \hline
    $H_0(v_3)$ &$5: v_3 \rightarrow v_1 \rightarrow v_2, ~6: v_3 \rightarrow v_2 \rightarrow v_4, ~7: v_3 \rightarrow v_1$\\
    \hline
    $H_0(v_4)$ & $8: v_4 \rightarrow v_3 \rightarrow v_1, ~9: v_4 \rightarrow v_3$\\
    \hline
    \multicolumn{2}{c}{(a) Index structure $H_0$}
\end{tabular}
}

\centering
{
\renewcommand{\arraystretch}{1.3}
\begin{tabular}{|c|l|}
\hline
    & record $c = \{id, step\} $\\
    \hline
    $C^V_0(v_1)$ & $\{1,0\},~\{2,0\},~\{5,1\}$\\
    \hline
    $C^V_0(v_2)$ & $\{1,1\},~\{3,0\},~\{4,0\},~\{6,1\}$\\
    \hline
    $C^V_0(v_3)$ & $\{4,2\},~\{5,0\},~\{6,0\},~\{7,0\},~\{8,1\}$ \\
    \hline
    $C^V_0(v_4)$ & $\{4,1\},~\{8,0\},~\{9,0\}$\\
    \hline
    \multicolumn{2}{c}{(b) Auxiliary structure $C^V_0$}
\end{tabular}
}

\centering
\caption{An example of the index and auxiliary structures.}
\label{exm:original}
\end{small}
\end{table}

Algorithm \ref{algo:ins} shows the pseudo-code of our {\bf Update-Insert} algorithm to handle index update with an edge insertion. The symbol $C^V$ denotes an auxiliary data structure of index $H$. Each element $c\mathrm{=}\{id, step\} \in C^V(u)$ is a crossing record, representing that the random walk with ID $c.id$ crosses $u$ at step $c.step$. We update $\past{H}$ to $\cur{H}$ after the insertion of $\cur{e}\mathrm{=}\tuple{\cur{u},\cur{v}}$ by {\em (i)} sampling records in $C^V(\cur{u})$ with probability $1\mathrm{/}\cur{d}(\cur{u})$; {\em (ii)} updating sampled random walks on the evolved graph. In particular, for each sampled record $c$ from $\past{C^V}(\cur{u})$, we first redirect the random walk with ID $c.id$ to the newly inserted edge $\cur{e}$ at its ($c.step\mathrm{+}1$)-th step (Line \ref{algo:ins:upd-redirected}), and then make it continue to randomly traverse until it has the same hops as the random walk with ID $c.id$ in $\past{H}$ (Line \ref{algo:ins:upd-continue}). After updating existing random walks, we further add additional random walks starting from $\cur{u}$ into $\cur{H}$ (Lines \ref{algo:ins:add}-\ref{algo:ins:add-end}) to increase the size of $\cur{H}(\cur{u})$ to fulfil conditions in Lemma \ref{lm:fora:index} and then finishes the update.

\begin{table}[t]
\begin{small}

\centering
{
\renewcommand{\arraystretch}{1.1}
\begin{tabular}{|c|l|}
    \hline
    $H_1(v_1)$ & $1: v_1 \rightarrow v_2 \rightarrow v_4,~2: v_1 \rightarrow v_2 $ \\
    \hline
    $H_1(v_2)$ & $3: v_2 \rightarrow v_4, ~4: v_2 \rightarrow v_4 \boldsymbol{\rightarrow v_1 \rightarrow v_2} $ \\
    \hline
     $H_1(v_3)$ &$5: v_3 \rightarrow v_1 \rightarrow v_2, ~6: v_3 \rightarrow v_2 \rightarrow v_4, ~7: v_3 \rightarrow v_1$\\
    \hline
    $H_1(v_4)$ & $8: v_4 \boldsymbol{\rightarrow v_1 \rightarrow v_2}, ~9: v_4 \rightarrow v_3, ~\underline{10: v_4 \rightarrow v_3}$ \\
    \hline
    \multicolumn{2}{c}{(a) The index structure $H_1$}
\end{tabular}
}

\centering
{
\renewcommand{\arraystretch}{1.3}
\begin{tabular}{|c|l|}
    \hline
    $C^V_1(v_1)$ & $\{1,0\},~\{2,0\},\boldsymbol{~\{4,2\}},~\{5,1\},~\boldsymbol{\{8,1\}}$\\
    \hline
    $C^V_1(v_2)$ & $\{1,1\},~\{3,0\},~\{4,0\},~\{6,1\}$\\
    \hline
    $C^V_1(v_3)$ & $\xcancel{\{4,2\}},~\{5,0\},~\{6,0\},~\{7,0\},~\xcancel{\{8,1\}}$\\
    \hline
    $C^V_1(v_4)$ & $\{4,1\},~\{8,0\},~\{9,0\},~\boldsymbol{\{10,0\}}$\\
    \hline
    \multicolumn{2}{c}{(b) The updated entries in auxiliary structure $C^V_1$}
\end{tabular}
}

\centering
\vspace{0mm}
\caption{Updated index and $C^V$ after Insertion.}
\label{exm:ins}

\end{small}
\end{table}

\begin{example}\label{exmp:ins}
Consider an original graph $G_0$ as shown in Fig.~\ref{exm:graph}(a), then assume that the initial index structure $H_0$ and auxiliary structures $C^V_0$ are as shown in Table \ref{exm:original}. In Table~\ref{exm:original}(a), the row labeled with $H_0(v_i)$ stores the random walks which start from $v_i$. For example, there are two random walks (which have IDs 1 and 2) starting from $v_1$. To simplify the description, we use $\mathrm{RW_i}$ as the representation of the random walk with ID $i$. In Table~\ref{exm:original}(b), the row labeled with $C^V_0(v_i)$ records which random walks visit node $v_i$. In the case as Table~\ref{exm:original} shows, there are three records: $\{1,0\},\{2,0\},\{5,1\}$ in the row corresponding to $v_1$. The record of $\{5,1\}$ represents that $\mathrm{RW_5}$ visits $v_1$ at its $1$st step, which can be verified by the detail of $\mathrm{RW_5}$ showed in Table~\ref{exm:original}(a). Now, when an edge insertion $\tuple{v_4,v_1}$ comes as shown in Fig.~\ref{exm:graph}(b), the index will be updated by following steps:

\begin{enumerate}[topsep=0mm, partopsep=0pt, itemsep=0pt, leftmargin=10pt,label = (\roman*)]
\item Sample random walks in $C^V_0(v_4)$ with probability $1\mathrm{/}2$ (as the out-degree of $v_4$ at timestamp $1$ is $2$) and then add sampled random walks to $C$. We assume that $\mathrm{RW_4}$ and $\mathrm{RW_8}$ are chosen to be updated as $C^V_0(v_4)$ contains records $\{4,1\},\{8,0\},\{9,0\}$.
\item Redirect $\mathrm{RW_4}$ to $v_1$ at its $2$nd step and re-sample from its $3$rd step, since the only out-neighbour of $v_1$ is $v_2$, $\mathrm{RW_4}$ will visit $v_2$ at step $3$. Then $\mathrm{RW_4}$ has a length of $3$ that is equal to the number of hops it held in $H_0$, so it will stop. A similar process is applied to $\mathrm{RW_8}$ as well. Table~\ref{exm:ins}(a) shows a possible repaired index where the adjusted part is shown in boldface.
\item Add one more random walk (underlined) to fulfill the requirement as the out-degree of $v_4$ is incremented by $1$. 
\item $C^V$ gets updated to reflect changes of $\mathrm{RW_4}$, $\mathrm{RW_8}$ and the new random walk $\mathrm{RW_{10}}$. For example, $\{8,1\}$ in $C^V_0(v_3)$ is moved into $C^V_1(v_1)$. Table~\ref{exm:ins}(b) shows the updated $C^V_1$, where new records are in boldface and obsoleted records are scratched. \done
\end{enumerate}

\end{example}

To show the correctness of {\bf Update-Insert}, we first make a review of FORA-like approaches in Section~\ref{sec:sol:static}. Such methods roughly approximate PPR scores in Forward-Push phase and then refine the estimation using the sampled random walks. Thus, we have:

\begin{lemma}
An index-based FORA-like solution gives a guarantee of $(\epsilon, \delta)$-approximation for the result if the following conditions hold:
\begin{itemize}[topsep=0mm, partopsep=0pt, itemsep=0pt, leftmargin=11pt]
    \item{\bf Independence.} The random walks in $H$ are independent.
    \item{\bf Adequateness.} For each node $s$, the set $H(s)$ of pre-stored random walks starting from $s$ satisfies Lemma~\ref{lm:fora:index}.
    \item{\bf Unbiasedness.} Let $H(s,t)$ denote the subset of $H(s)$ where $w \in H(s,t)$ means $w$ starting from $s$ terminates at $t$. For arbitrary nodes $s,t \in V$, $|H(s,t)|\mathrm{/}|H(s)|$ is an unbiased estimation of $\pi(s,t)$.
\end{itemize}
\end{lemma}
\begin{proofsketch}
The adequateness condition directly comes from Lemma~\ref{lm:fora:index}, and Lemma~\ref{lm:fora:index} is derived from a concentration bound (more precisely, Theorem 5 in~\cite{ChungLu06}), of which the independence and unbiasedness conditions are preconditions. We need to meet these conditions so that the accuracy of PPR scores can be guaranteed.
\end{proofsketch}

Hence, if we build an initial index $H_0$ which is directly sampled independently on $G_0$ that contains a sufficient number of random walks from each node to satisfy the above conditions, the correctness of {\bf Update-Insert} can be derived from the following result:

\vspace{-0.5mm}
\begin{theorem}\label{thm:insert}
$\cur{H}$ provided by Algorithm~\ref{algo:ins} satisfies the independence, adequateness, and unbiasedness if $\past{H}$ holds these properties.
\end{theorem}

\vspace{-0.5mm}
The above results indicate that the updated index still satisfies three conditions and provides the approximation guarantee. Notably, the update cost to the random walk index is very light and can be bounded with $O(1)$ in expectation as shown in Theorem \ref{thm:insert-cost}. 

\vspace{-0.5mm}
\begin{theorem}\label{thm:insert-cost}
Given an inserted edge that follows the random arrival model (Definition \ref{def:ram}), Algorithm \ref{algo:ins} takes $O(1)$ expected time to update the index when $\rmax\mathrm{\cdot} \omega \mathrm{=} \Theta(1)$.
\end{theorem}

\vspace{-0.5mm}
For ease of exposition, all proofs are deferred to Section \ref{sec:analysis}.

\subsection{Edge Deletion}\label{sec:iprea:del}
\begin{algorithm}[t]
\caption{Update-Delete}\label{algo:del}
\LinesNumbered
\KwIn{Graph $\cur{G} = (\cur{V}, \cur{E})$, past index $\past{H}$, edge $\cur{e} = \tuple{\cur{u},\cur{v}}$.}
\KwOut{New index $\cur{H}$.}
  $\cur{H} \leftarrow \past{H}$\;
  $W^* \leftarrow \emptyset$\;
  \While {$|\cur{H}(\cur{u})| > \mceil{\cur{d}(\cur{u}) \cdot \rmax \cdot \omega}$}{ \label{algo:del:rm}
    $w^* \leftarrow \funcname{Sample-Uniform}(\cur{H}(\cur{u}))$\;
    $\cur{H} \leftarrow \cur{H} \setminus \{w^*\}$\;
    $W^* \leftarrow W^* \cup \{w^*\}$
  }\label{algo:del:rm-end}
  \ForEach {$c \in \past{C^E}(\cur{e}) \setminus \past{C^E}(W^*)$} {\label{algo:del:upd}
    $w \leftarrow \cur{H}[c.id]$\;
    $\cur{H}[c.id] \leftarrow \funcname{Walk-Restart}(G',w,c.step)\}$\;
  }\label{algo:del:upd-end}
\Return{$\cur{H}$}\;
\end{algorithm}

When an edge deletion to $\tuple{\cur{u},\cur{v}}$ occurs, the main idea to update the random walks with an edge deletion is similar to that of edge insertion, by updating the affected random walks that cross $\cur{u}$. 

Algorithm \ref{algo:del} shows the pseudo-code of {\bf Update-Delete}. We use the symbol $C^E$ to denote the auxiliary structure of $H$ where each $c \in C^E(e)$ is a crossing record where $c\mathrm{=}\{id, step\}$ denotes the random walk $c.id$ passes through $e$ between step $c.step$ and $c.step\mathrm{+}1$. Note that we require a smaller number of random walks starting from $\cur{u}$ after the deletion of $\cur{e}\mathrm{=}\tuple{\cur{u},\cur{v}}$. That means some random walks in $\past{H}(\cur{u})$ are no longer needed. To avoid the waste of space to store the index, we uniformly select some random walks $w^* \in \past{H}(u)$ and remove them before adjusting the index (Lines \ref{algo:del:rm}-\ref{algo:del:rm-end}). Then, for each record $c\in \past{C^E}(\cur{e})$ (except records corresponding to recently trimmed random walks, denoted as $\past{C^E}(W^*)$), the corresponding random walk $c.id$ must be repaired since $\cur{e}$ no longer exists in $\cur{G}$. Let such a random walk restart at step $c.step$ where it hits node $\cur{u}$ before passing through edge $\cur{e}$ (Lines \ref{algo:del:upd}-\ref{algo:del:upd-end}). Then, it randomly traverses on $\cur{G}$ until it has the same hops as the $c.id$ random walk in $\past{H}$. For other random walks, we keep them as they are in $\past{H}$.

\begin{table}
\begin{small} 

\centering
{
\renewcommand{\arraystretch}{1.3}
\begin{tabular}{|c|l|c|l|}
    \hline
    $C^E_1(v_1,v_2)$ & \multicolumn{3}{l|}{ $\{1,0\},~\{2,0\},~\{4,2\},~\{5,1\},~\{8,1\}$} \\
    \hline
    $C^E_1(v_2,v_4)$ & \multicolumn{3}{l|}{ $\{1,1\},~\{3,0\},~\{4,0\},~\{6,1\}$} \\
    \hline
    $C^E_1(v_3,v_1)$ & $\{5,0\},~\{7,0\}$ & $C^E_1(v_3, v_2)$ & $\{6,0\}$ \\
    \hline
    $C^E_1(v_4, v_1)$ & $\{4, 1\},~\{8, 0\}$ &$C^E_1(v_4, v_3)$ & $\{9, 0\},~\{10, 0\}$ \\
    \hline
    \multicolumn{4}{c}{(a) Auxiliary Structure $C^E_1$}
\end{tabular}
} 

\centering
{
\renewcommand{\arraystretch}{1.1}
\begin{tabular}{|c|l|}
    \hline
    $H_2(v_3)$ & $5: v_3 \rightarrow v_1 \rightarrow v_2, ~6: v_3 \boldsymbol{\rightarrow v_1 \rightarrow v_2}, ~\xcancel{7: v_3 \rightarrow v_1}$\\
    \hline
    \multicolumn{2}{c}{(b) The updated index entries in $H_2$}
\end{tabular}
}

\centering
{
\renewcommand{\arraystretch}{1.3}
\begin{tabular}{|c|l|c|l|}
\hline
    $C^E_2(v_1,v_2)$ & \multicolumn{3}{l|}{$\{1,0\},~\{2,0\},~\{4,2\},~\{5,1\},~\boldsymbol{\{6,1\}},~\{8,1\}$}\\
    \hline
    $C^E_2(v_2,v_4)$ &\multicolumn{3}{l|}{ $\{1,1\}, ~\{3,0\}, ~\{4,0\},~\xcancel{\{6,1\}}$}\\
    \hline
    $C^E_2(v_3,v_1)$ & $\{5,0\}, ~\boldsymbol{\{6,0\}},~\xcancel{\{7,0\}}$ & $\xcancel{C_2^E(v_3, v_2)}$&$\xcancel{\{6,0\}}$  \\
    \hline
    $C^E_1(v_4, v_1)$ & $\{4, 1\},~\{8, 0\}$ &$C^E_1(v_4, v_3)$ & $\{9, 0\},~\{10, 0\}$ \\
    \hline
    \multicolumn{4}{c}{(c) The updated entries in auxiliary structure $C^E_2$}
\end{tabular}
} 

\centering
\vspace{0mm}
\caption{Index and Auxiliary Structures after Deletion.}
\label{exm:del}
\end{small}
\end{table}

\begin{example}\label{exmp:del}
Assume that the current graph is $G_1$ as shown in Figure~\ref{exm:graph}(b) after the insertion in Example~\ref{exmp:ins}. Then, when the deletion of edge $\tuple{v_3,v_2}$ comes, it will evolve to $G_2$ as shown in Figure~\ref{exm:graph}(c). To simplify the description, we omit the auxiliary structure $C^V$ on nodes as it is not used in deletion and it will be maintained in the same way as we described in Example~\ref{exmp:ins}. After the edge deletion, we will update the index with the following steps:
\begin{enumerate}[topsep=0mm, partopsep=0pt, itemsep=0pt, leftmargin=1.5em,label = (\roman*)]
\item Remove one random walk in $H_2(v_3)$ to save space. We are allowed to do it as the out-degree of $v_3$ is decreased by $1$. Assume that $\mathrm{RW_7}$ is removed (crossed-out as shown in Table~\ref{exm:del}(a)).
\item Repair $\mathrm{RW_6}$ since it passes through edge $\tuple{v_3,v_2}$ between its step $0$ and step $1$. We simulate the restarting part of $\mathrm{RW_6}$ from its step $0$. A possible repaired index is as shown in Figure~\ref{exm:del}(b) where the newly removed random walk is crossed-out and the repaired parts of existing random walks are marked in boldface. The other entries $H(\cdot)$ that are not changed are omitted.
\item To match the updated random walks, the auxiliary structure $C^E$ needs to be updated accordingly. For example, since $\mathrm{RW_6}$ is changed from $v_3 \mathrm{\rightarrow} v_2  \mathrm{\rightarrow} v_4$ to $v_3 \mathrm{\rightarrow} v_1 \mathrm{\rightarrow} v_2$, the records $\{6,0\}$ and $\{6,1\}$ are moved into $C^E_2(v_3,v_1)$ and $C^E_2(v_1,v_2)$ from $C^E_1(v_3, v_2)$ and $C^E_1(v_2, v_4)$, respectively. Table \ref{exm:del}(c) shows the updated $C^E_2$ where the new records are boldfaced and the obsoleted records are scratched. Note that we also scratch the entry $C^E_2(v_3,v_2)$ in Table \ref{exm:del}(c) as it is stale due to the deletion of edge $\tuple{v_3,v_2}$.  \done
\end{enumerate}
\end{example}

Similar as we described in Section~\ref{sec:iprea:ins}, the following result gives the correctness of {\bf Update-Delete} since we can build an initial index $H_0$ satisfying independence, adequateness, and unbiasedness.

\vspace{-0.5mm}
\begin{theorem}\label{thm:delete}
$\cur{H}$ provided by Algorithm~\ref{algo:del} satisfies the independence, adequateness, and unbiasedness if $\past{H}$ holds these properties.
\end{theorem}

\vspace{-0.5mm}
Just like the edge insertion case, the update cost will be light if an edge deletion follows the random arrival model. As Theorem~\ref{thm:delete-cost} shows, the cost can be bounded by $O(1)$.

\vspace{-0.5mm}
\begin{theorem}\label{thm:delete-cost}
Given an edge deletion that follows the random arrival model (Definition \ref{def:ram}), Algorithm \ref{algo:ins} takes $O(1)$ expected time to update the index when $\rmax\mathrm{\cdot} \omega \mathrm{=} \Theta(1)$.
\end{theorem}

\vspace{-0.5mm}
For ease of exposition, all proofs are deferred to Section \ref{sec:analysis}.

\subsection{A New Sampling Scheme for Index Update} \label{sec:sample}
Up to now, we have shown the general framework of our solution which updates the index within constant time for each edge insertion/deletion. However, a simple implementation of Algorithm~\ref{algo:ins} and Algorithm~\ref{algo:del} will result in several times space  over existing solutions, if we maintain auxiliary structures in views of both nodes and edges. Next, we show a new sampling scheme for index updates to reduce the space consumption of auxiliary structures.

\header
{\bf Main Challenge.}
As we described in Sections \ref{sec:iprea:ins} and \ref{sec:iprea:del}, our solution achieves superb update efficiency since the expected number of random walks affected by an edge update is $O(1)$. However, it means that we need to trace each affected random walk within $O(1)$ time as well. To achieve this, we maintain auxiliary structures in the view of nodes (denoted as $C^V$) in Algorithm~\ref{algo:ins} and in the view of edges (denoted as $C^E$) in Algorithm~\ref{algo:del}, and thus our solution requires several times space to save the index. Therefore, it deserves our effort to use only one auxiliary structure to support both sampling among random walks crossing a node $u$ and finding all random walks passing through an updated edge $e\mathrm{=}\tuple{u,v}$.

Since we need to exactly find all random walks passing through the edge $\cur{e}$ in Algorithm~\ref{algo:del}, it is difficult to use only $C^V$ to achieve our goal. Thus, we turn to find an alternative sampling scheme for index update to sample random walks using $C^E$ in Algorithm~\ref{algo:ins}.

Before we introduce how to sample with $C^E$ in the insertion case, there is an issue that a random walk starting from $s$ maybe terminate directly with probability $\alpha$ so that it does not pass through any edge and therefore cannot be recorded by any edge. To fix the problem, consider a variant of Equation \ref{eqn:fora:fpinvar} derived in \cite{WYWXWLYT19}:
\begin{flalign*}
  \pi(s,t) &= \pih(s,t) + \sum_{v \in V} r(s,v) \cdot (\sum_{l=0}^{\infty}\pi^l(v,t)) \\
  &= \pih(s,t) + \sum_{v \in V} r(s,v) \cdot (\pi^0(v,t)+\pi^+(v,t)), 
\end{flalign*}
where $\pi^l(v,t)$ is the $l$-hop PPR which denotes the random walk starting from $v$ stops at $t$ exactly at its $l$-th step, and $\pi^+(v,t)$ is the sum of $\pi^l(v,t)$ for all non-zero $l$. Note that, $\pi^0(v,t)\mathrm{=}\alpha$ only when $v\mathrm{=}t$, and $\pi^0(v,t)\mathrm{=}0$ otherwise. Because of the certainty of $\pi^0(v,\cdot)$, we need not store the random walks which terminate at their source node immediately, and thus $C^E(e)$ is complete to tracing all random walks we need to maintain.

Recall that after the edge $\cur{e}\mathrm{=}\tuple{\cur{u}, \cur{v}}$ is inserted, we need to sample each record crossing $\cur{u}$ with probability $1\mathrm{/}\cur{d}(\cur{u})$. For the convenience of description, we will use the symbol $C^V$ and $C^E$ which have no subscript to represent $\past{C^V}$ and $\past{C^E}$ respectively, and then let $c^V(u)$ and $c^E(e)$ denote the size of $C^V(u)$ and $C^E(e)$ respectively. The naive way is to generate random numbers for every record in $C^V(\cur{u})$ to decide whether a record will be adjusted or not. This approach can be easily implemented with $C^E$ where we can access the records in each $C^E(e)$ for $e\in \past{\Edge}({\cur{u}})$ and then roll the dice. However, such a method uses $O(c^V(\cur{u}))$ time which will be $O(\past{d}(\cur{u}))\mathrm{=}O(\cur{d}(\cur{u}))$ in expectation to select the random walks to be adjusted while only $O(1)$ random walks will be selected as we claimed in Section~\ref{sec:iprea:ins}.

An alternative approach is geometric sampling. We number the records in $C^V(\cur{u})$ from $1$ to $c^V(\cur{u})$, and use an iterator to indicate which record we are visiting. Geometric sampling works as follows: assume the iterator currently points to $i$ (initially $i\mathrm{=}0$), it generates a random number $j$ from geometric distribution $Geom(1\mathrm{/}\cur{d}(\cur{u}))$, then it {\em (i)} stops if $i\mathrm{+}j\mathrm{>}c^V(\cur{u})$; or {\em (ii)} jumps to $i\mathrm{+}j$ and selects the pointed record. It is proved that this approach can sample each record with probability $1\mathrm{/}\cur{d}(\cur{u})$, and since the iterator has only visited the records which will be selected, the expected time for sampling will be $O(1)$. Nonetheless, it is hard to apply the approach to $C^E$ efficiently. To explain, if we only maintain the auxiliary structure $C^E$ on edges but not $C^V$ on nodes, we need to find all random walks that pass through each edge $e\in\past{\Edge}(\cur{u})$ and then apply the geometric sampling approach to select the random walks to be adjusted. However, by examining each out-going edge of $u$, it already takes $O(\cur{d}(\cur{u}))$ time. We may make use of advanced data structures (e.g. segment-tree) to accelerate the procedure of seeking for the selected random walk among $C^E$. However, searching on a tree structure makes a factor of $O(\log{\cur{d}(\cur{u})})$ increment of time complexity. More importantly, our motivation to use one auxiliary structure to support the tracing for both edge insertion and deletion is to reduce the space consumption of the auxiliary structure but the size of such an auxiliary structure of auxiliary structure to handle the whole graph will also be $O(m)$.

\header
{\bf Our Solution.} To tackle this issue, we first change the above geometric sampling approach to a sampling approach via the binomial distribution. In particular, we sample the number of random walk records (crossing $u$) from binomial distribution $B(c^V(\cur{u}), 1\mathrm{/}\cur{d}(\cur{u}))$, and then sample the records in $C^V(\cur{u})$ without replacement. It is easy to verify that this approach is equivalent to geometric sampling. However, the binomial sampling approach is still difficult to apply to $C^E$ directly while meeting both time and space efficiency for the same reason as the geometric sampling approach.

To further overcome the obstacle, we present the following idea. In the process of binomial sampling, we repeat to sample a record from $C^V(\cur{u})$ without replacement. To avoid an iteration among $\past{\Edge}({\cur{u}})$ (which takes $O(d(\cur{u}))$ time), we can first sample an edge $e \in \past{\Edge}({\cur{u}})$ which has at least one record (called an active edge) and then uniformly select one of the records $c \in  C^E(e)$. It is obvious that if we select the edge $e$ with probability $c^E(e)\mathrm{/}c^V(\cur{u})$, we reproduce the uniformly sampling in $C^V(\cur{u})$. However, a weighted random sampling within $O(1)$ time requires the help of advanced data structures (e.g. alias array) with total size $O(m)$. Again, we will not build any additional data structure which has a total size of $O(m)$ since it goes against our intention to reduce space overheads. Instead, we turn to sample an edge $e \in \past{\Edge}({\cur{u}})$ with probability $1\mathrm{/}d'(\cur{u})$, where $d'(\cur{u})$ is the number of active edges and this number can be easily maintained with a counter for each node, saving far more space than maintaining auxiliary structure $C^V$ on nodes.

To check if the alternative binomial sampling also keeps the unbiasedness, it needs to be clarified that we will prove the unbiasedness of $\cur{H}$ in Section~\ref{sec:analysis:corr} only based on the fraction of the number of selected records in $C^E(e)$. Thus, any records passing through the same edge $e$ can be equivalent for providing unbiasedness. Hence, if we select a record $c \in C^E(e)$, uniformly selecting another record $c'\in C^E(e)$ instead will not affect the unbiasedness. Assume that $c^V(\cur{u})\mathrm{=}\mathscr{w}$ and $d'(\cur{u})\mathrm{=}\mathscr{d}$, consider the probability that the record selected by the original binomial sampling passes through edge $e$:
\begin{flalign}
    \Pr[c\in C^E(e)] &= \sum_{k=1}^{\mathscr{w}}\Pr[c \in C^E(e)|c^E(e) = k] \cdot \Pr[c^E(e) = k] \notag \\
    &= \sum_{k=1}^{\mathscr{w}}\frac{k}{\mathscr{w}}\cdot \binom{\mathscr{w}}{k}\left(\frac{1}{\mathscr{d}}\right)^k\left(1-\frac{1}{\mathscr{d}}\right)^{\mathscr{w}-k}. \label{eqn:edge-sample}
\end{flalign}
Firstly, we have that:
\begin{flalign*}
    \frac{k}{\mathscr{w}}\cdot \binom{\mathscr{w}}{k} &= \frac{k}{\mathscr{w}} \cdot \frac{\mathscr{w}!}{k!\cdot(\mathscr{w}-k)!} 
    = \frac{(\mathscr{w}-1)!}{(k-1)!\cdot(\mathscr{w}-k)!} = \binom{\mathscr{w}-1}{k-1} ,
\end{flalign*}
and thus Equation \ref{eqn:edge-sample} will be:
\begin{flalign*}
     \Pr[c\in  C^E(e)] &= \sum_{k=1}^{\mathscr{w}}\binom{\mathscr{w}-1}{k-1} \left(\frac{1}{\mathscr{d}}\right)^k\left(1-\frac{1}{\mathscr{d}}\right)^{\mathscr{w}-k} \\
    &= \frac{1}{\mathscr{d}}\cdot\sum_{k=0}^{\mathscr{w}-1} \binom{\mathscr{w}-1}{k} \left(\frac{1}{\mathscr{d}}\right)^k\left(1-\frac{1}{\mathscr{d}}\right)^{\mathscr{w}-1-k} = \frac{1}{\mathscr{d}} \; ,
\end{flalign*}
where the last equality is due to the fact that the cumulative term is the summation of the probability mass function of $B(\mathscr{w}\mathrm{-}1, 1\mathrm{/}\mathscr{d})$ and thus equals to $1$. Therefore, if $\past{H}$ satisfies the unbiasedness, for any fixed $c^V(\cur{u})$ and $d'(\cur{u})$, the original binomial sampling method samples a record from a particular edge $e$ with probability $1\mathrm{/}d'(\cur{u})$, which is consistent with the strategy which samples the edges of $\cur{u}$ uniformly. Since the unbiasedness of $\past{H}$ implies that the number of records crossing $u$ and the number of active out-going edges of $u$ are also unbiased, we can see that our strategy is equivalent to the original binomial sampling to keep the unbiasedness of $\cur{H}$.

\begin{algorithm}[t]
\caption{Edge-Sampling}\label{algo:new-sampling}
\LinesNumbered
\KwIn{Node $u$, record lists on edges $C^E$.}
\KwOut{Sampled records $C$.}
$C \leftarrow \emptyset$\;
$k \sim B(\cur{\mathscr{c}}(\cur{u}), \frac{1}{\cur{d}(\cur{u})})$\;
\For{$i$ from $1$  to $k$} {
    $\Edge \leftarrow \funcname{Active-Edges}(\cur{u})$\;
    $e \leftarrow \funcname{Sample-Uniform}(\Edge)$\;
    $c \leftarrow \funcname{Sample-Uniform}(C^E(e))$\;
    $C \leftarrow C \cup \{c\}$\;
}
\Return{$C$}\;
\end{algorithm}

So far, we have the final solution to sample records among $C^E(e)$. Algorithm \ref{algo:new-sampling} shows the pseudo-code of the new sampling method, where we use $\mathscr{c}(u)$ to denote the total number of records crossing $u$. Note that $\mathscr{c}(u)$ is equal to $c^V(u)$ which is the size of $C^V(u)$ but we store $\mathscr{c}(u)$ as a counter directly on each node, which takes $\Theta(n)$ space and is much smaller than $O(m)$ in practice. Now return to Algorithm \ref{algo:new-sampling}, it first generates the number of records to be selected from the binomial distribution $B(\cur{\mathscr{c}}(\cur{u}), 1\mathrm{/}\cur{d}(\cur{u}))$ (Line 2). To sample each record, it uniformly samples an edge $e\in \past{\Edge}({\cur{u}})$ which is active and then uniformly samples one record in $c^E(e)$ (Lines 4-7). Finally, it returns the sampled records. We apply Algorithm~\ref{algo:new-sampling} to Algorithm~\ref{algo:ins} instead of sampling directly in $C^V(\cur{u})$ (Line \ref{algo:ins:upd}) to select the affected random walks. For other parts, e.g., Lines \ref{algo:del:rm}-\ref{algo:del:rm-end} in Algorithm \ref{algo:del}, the sampling approaches are unchanged.

\section{Theoretical Analysis}\label{sec:analysis}
\subsection{Proofs of Correctness}\label{sec:analysis:corr}
\header{\bf Proof of Theorem \ref{thm:insert}.} Firstly, Algorithm~\ref{algo:ins} samples in $\past{C^V}(\cur{u})$ and updates the sampled random walks independently. In addition, for the newly sampled random walks (Lines 9-10), they are sampled independently and are independent from other random walks in $\cur{H}$. Thus, $\cur{H}$ satisfies the independence property. Then, since the only difference between $\cur{G}$ and $\past{G}$ is the edge $\cur{e}\mathrm{=}\tuple{\cur{u}, \cur{v}}$, we add additional random walks starting from $\cur{u}$ to satisfy the adequateness (Lines 9-10). Next, we focus on unbiasedness.

It is clear that the additional random walks are unbiased, thus we will omit the clarification of them below. Consider the random walks in $\past{H}$ which never cross $\cur{u}$. Let $\pi(s,t;\bar{u})$ denote the contribution to $\pi(s,t)$ without crossing node $u$, and the following relation holds:
\begin{equation}
    \cur{\pi}(s,t;\cur{\bar{u}}) = \past{\pi}(s,t;\cur{\bar{u}}), \label{eqn:upd:invar}
\end{equation}
\noindent because the insertion of $\cur{e}$ makes no effect on the random walks which never cross $\cur{u}$. Since $\past{H}$ is unbiased and Algorithm~\ref{algo:ins} makes no change to the random walks which never cross $\cur{u}$ in $H(s)$ (denoted as $H(s;\bar{u})$, correspondingly, the crossing subset denoted as $H(s;u)$), we obtain that,
\begin{equation}
\E\left[\frac{\cur{H}(s,t; \cur{\bar{u}})}{\cur{H}(s)}\right] = \cur{\pi}(s,t;\cur{\bar{u}}). \label{eqn:upd:enoxos}
\end{equation}

For the other part of random walks which still cross $\cur{u}$, let $\pi(s,t;e)$ denotes the contribution to $\pi(s,t)$ with a passing through of edge $e\mathrm{=}\tuple{\cur{u},v}$. To simplify the discussion, for the present, we assume any random walk will not cross $\cur{u}$ more than once and will discuss how to deal with the case that crosses $u$ multiple times later. Due to the memorylessness of the random walk process, we have:
\begin{flalign}
    \cur{\pi}(s,t;e) &= \Pr[w \in \cur{T}(e|s)] \cdot \cur{\pi}(v,t;\cur{\bar{u}}) \notag \\ &= \frac{1}{\cur{d}(\cur{u})} \cdot \Pr[w \in \cur{T}(\cur{u}|s)] \cdot \cur{\pi}(v,t;\cur{\bar{u}}), \label{eqn:ppr:comb}
\end{flalign}
\noindent where $w \in T(e|s)$ indicates the $s$-starting random walk $w$ passes through edge $e$ and $w \in T(u|s)$ indicates such a random walk crosses node $u$. Notably:
\begin{equation}
\Pr[w \in \cur{T}(\cur{u}|s)] = \Pr[w \in \past{T}(\cur{u}|s)] \label{eqn:upd:invar2}
\end{equation}
\noindent since a random walk must cross $\cur{u}$ before decides to pass through $\cur{e}$ or not. Combining Equations \ref{eqn:upd:invar},\ref{eqn:ppr:comb},\ref{eqn:upd:invar2} and since $\cur{d}(\cur{u})\mathrm{=}\past{d}(\cur{u})\mathrm{+}1$ with the insertion of $\cur{e}$, for $e\mathrm{=}\tuple{\cur{u},v} \in \past{\Edge}(\cur{u})$, we have:
\begin{flalign}
    \cur{\pi}(s,t;e) &= \frac{\past{d}(\cur{u})}{\cur{d}(\cur{u})}\cdot \frac{1}{\past{d}(\cur{u})} \cdot \Pr[w \in \past{T}(\cur{u}|s)] \cdot \past{\pi}(v,t;\cur{\bar{u}}) \notag \\
    &= \left(1\mathrm{-}\frac{1}{\cur{d}(\cur{u})}\right)\cdot \past{\pi}(s,t;e), \label{eqn:upd:ins1}
\end{flalign}

Lastly and clearly, the contribution of the inserted edge $\cur{e}$ is:
\begin{flalign}
    \cur{\pi}(s,t;\cur{e}) &= \frac{1}{\cur{d}(\cur{u})}\cdot \Pr[w \in \cur{T}(\cur{u}|s)] \cdot \cur{\pi}(v,t;\cur{\bar{u}}) \notag \\
    &= \frac{1}{\cur{d}(\cur{u})}\cdot \Pr[w \in \past{T}(\cur{u}|s)] \cdot \cur{\pi}(v,t;\cur{\bar{u}}). \label{eqn:upd:ins2}
\end{flalign}

In Algorithm~\ref{algo:ins}, we sample each $\cur{u}$-crossing record with probability $1\mathrm{/}\cur{d}(\cur{u})$ thus making the $1\mathrm{/}\cur{d}(\cur{u})$ fraction of descent to the contribution of random walks passing through $e \in \past{\Edge}(\cur{u})$ in expectation which keeps pace with Equation \ref{eqn:upd:ins1}. Then we force the sampled random walks to cross the edge $\cur{e}$ and continue randomly walking on $\cur{G}$ which exactly matches Equation~\ref{eqn:upd:ins2}. Due to the unbiasedness of $\past{H}$, the sum among $\cur{\Edge}(\cur{u})$ leads to:
\begin{equation}
\E\left[\frac{\cur{H}(s,t; \cur{u})}{\cur{H}(s)}\right] = \cur{\pi}(s,t;\cur{u}). \label{eqn:ins:exos}
\end{equation}

Equations \ref{eqn:upd:enoxos} and \ref{eqn:ins:exos} imply that $\cur{H}$ is unbiased if a random walk will not cross node $\cur{u}$ more than once. To show that $\cur{H}$ is still unbiased even if a random walk can cross $\cur{u}$ multi-times, let us consider a particular random walk in $\cur{H}$. Algorithm~\ref{algo:ins} independently samples each of $\cur{u}$-crossing records with probability $\cur{d}(u)$. Once a record $c$ is sampled, the rest path of the random walk $c.id$ after step $c.step$ will be discarded. Thus if multiple records of the same random walk are sampled, only the earliest one in the crossing order takes effect finally. Since the first time $i$ a random walk will turn to $\cur{e}$ after it crosses $\cur{u}$ obey a geometric distribution $Geom(1\mathrm{/}\cur{d}(\cur{u}))$ which consists of the probability that the $i$-th crossing record of it will be the dominant item in Algorithm~\ref{algo:ins}, we finish the proof. \done

\vspace{-2mm}
\header{\bf Proof of Theorem \ref{thm:delete}.} Algorithm~\ref{algo:del} first trims $\past{H}$ by uniformly sampling random walks from $\past{H}(\cur{u})$. Let $\past{H'}$ denote the trimmed $\past{H}$, and the independence and unbiasedness of $\past{H'}$ will be kept the same as $\past{H}$ for the property of uniform sampling. Then, the trimming stops when the size of $\past{H'}$ satisfies the adequateness exactly. Next, we repair the invalid random walks which pass through the recently deleted edge $\cur{e}$. According to Equation \ref{eqn:upd:enoxos} and since we make no change to the random walks without crossing $\cur{u}$, we have that Equation \ref{eqn:upd:invar} also holds for the deletion case.

For the other part of random walks, the deletion of $\cur{e}$ makes the contribution of $\cur{e}$ become $0$ and leads to an increment in the contribution of other edges $e \in \cur{\Edge}(\cur{u})$. By Equations \ref{eqn:upd:invar},\ref{eqn:ppr:comb},\ref{eqn:upd:invar2} and since $\cur{d}(\cur{u})\mathrm{=}\past{d}(\cur{u})\mathrm{-}1$ with the deletion of $\cur{e}$, we can see:
\begin{flalign*}
    \cur{\pi}(s,t;e) &= \left(1\mathrm{+}\frac{1}{\cur{d}(\cur{u})}\right)\cdot \frac{1}{\past{d}(\cur{u})} \cdot \Pr[w \in \cur{T}(\cur{u}|s)] \cdot \cur{\pi}(v,t;\cur{\bar{u}}) \\
    &=\past{\pi}(s,t;e) + \frac{1}{\past{d}(\cur{u})} \cdot \cur{\pi}(s,t;e),
\end{flalign*}
holds with the assumption that any random walk will not cross node $\cur{u}$ more than once. In Algorithm~\ref{algo:del}, we repair a random walk passing through $\cur{e}$ by restarting the random walk at $\cur{u}$, thus:
\begin{flalign*}
\E\left[\frac{\diffsym \cur{H}(s,t;e)}{\cur{H}(s)}\right] &= \E\left[\frac{\past{H}(s;\cur{e})}{\past{H}(s)}\right] \cdot \frac{1}{\cur{d}(\cur{u})} \cdot \cur{\pi}(v,t;\cur{\bar{u}}) \\
&= \frac{1}{\past{d}(\cur{u})\cur{d}(\cur{u})} \cdot \Pr[w \in \past{T}(\cur{u}|s)] \cdot \cur{\pi}(v,t;\cur{\bar{u}})\\
&= \frac{1}{\past{d}(\cur{u})} \cdot \cur{\pi}(s,t;e),
\end{flalign*}
where $\diffsym \cur{H}(s,t;e)$ denotes $\cur{H}(s,t;e)\mathrm{-}\past{H}(s,t;e)$ which is the increment of $H(s,t;e)$ at timestamp $\tau$, the second equality is based on the unbiasedness of $\past{H}$ and the third equality comes from Equations \ref{eqn:ppr:comb}-\ref{eqn:upd:invar2}. As the result of the above equations, we have:
\begin{equation}
\E\left[\frac{\cur{H}(s,t; e)}{\cur{H}(s)}\right] = \cur{\pi}(s,t;e). \label{eqn:del:exos}
\end{equation}

Thus, Equations \ref{eqn:upd:enoxos} and \ref{eqn:del:exos} imply that $\cur{H}$ is unbiased if a random walk will not cross $\cur{u}$ more than once. Finally, the correctness of Algorithm~\ref{algo:del} comes from the fact that the deletion of $\cur{e}$ makes no effect on a random walk unless it passes through $\cur{e}$, and the rest part of such a random walk behind the first time it passes through $\cur{e}$ is entirely stale in the deletion case. \done

\vspace{-2mm}
\subsection{Proofs of Efficiency}\label{sec:analysis:perf}
\header{\bf Proof of Theorem \ref{thm:insert-cost}.} As Algorithm~\ref{algo:ins} shows, the cost of {\bf Update-Insert} contains two parts. One is caused by the modification of random walks whose records are sampled in set $\past{C^V}(\cur{u})$ to reflect the change of out-neighbors of $\cur{u}$ due to the arrival of edge $\cur{e}$. Let $\cur{c}$ denote the number of random walks, we have:
\begin{equation*}
  \E\left[\cur{c}\right] \le \sum_{u \in \past{V}} \E\left[|\past{C^V}(u)|\right] \cdot \frac{1}{\cur{d}(u)} \cdot \Pr\left[ \cur{e} \in \cur{\Edge}(u) \right],
\end{equation*}
where the inequality is because we will update a random walk at most once even if multiple of its records are sampled. Then with Equation~\ref{eqn:pr:ins}, it yields that:
\begin{flalign*}
  \E\left[\cur{c}\right] &\le \sum_{u \in \past{V}} \E\left[|\past{C^V}(u)|\right] \cdot \frac{1}{\cur{d}(u)} \cdot \frac{\cur{d}(u)}{\cur{m}} \\
  &= \frac{1}{\cur{m}} \cdot \sum_{u \in \past{V}} \E\left[|\past{C^V}(u)|\right].
\end{flalign*}

In addition, according to Lemma~\ref{lm:fora:index}, the number of random walks starting from node $s$ should be $\mceil{d(s) \mathrm{\cdot} \rmax \mathrm{\cdot} \omega}$. Let $\past{h}(s,u)$ be the expected times that an $s$-starting random walk crosses node $u$ at timestamp $\taume$. It holds that: 
\begin{equation}\label{eqn:relation-ppr}
  \past{h}(s,u) = \frac{\emalpha}{\alpha} \cdot \past{\pi}(s,u),
\end{equation}
\noindent where $\past{\pi}(s,u)$ is the PPR score of node $u$ with respect to $s$ on graph $\past{G}$. To explain, $\past{\pi}(s,u)$ can be written as follows:
\begin{equation*}
    \past{\pi}(s,u) = \sum_{i=0}^\infty\past{\pi}^i(s,u),
\end{equation*}
where $\past{\pi}^i(s,u)$ is the probability that an $s$-starting random walk stops at $u$ exactly at the $i$-th hop. Therefore, the probability that an $s$-starting random walk visits node $u$ at the $i$-th hop can be written as $\past{\pi}^i(s,u)/\alpha$. Multiplying it by $\emalpha$, we derive the probability that an $s$-starting random walk crosses $u$ at the $i$-th hop is $\past{\pi}^i(s,u) \mathrm{\cdot}(\emalpha)\mathrm{/}\alpha$. Summing all hops together, we derive that $\past{h}(s,u)$, the expected times an $s$-starting random walk crossing $u$, satisfies Equation \ref{eqn:relation-ppr}. Then, we could express $|\past{C^V}(u)|$ as:
\begin{equation}\label{eqn:relation-ppr-affected-edge}
\E\left[|\past{C^V}(u)\right] = \sum_{s \in \past{V}} \mceil{\past{d}(s) \cdot \rmax \cdot \omega} \cdot \past{h}(s,u). 
\end{equation}

Let $\past{V}^*$ denote the set of nodes with at least one out-going edge, and let $\past{n}^* \mathrm{=} |\past{V}^*|$. Combining the above equations, we obtain:
\begin{flalign*}
    \E\left[ \cur{c} \right] &\le \frac{1}{\cur{m}} \cdot \sum_{u \in \past{V}} \sum_{s \in \past{V^*}} \mceil{ \past{d}(s) \cdot \rmax \cdot \omega } \cdot \frac{\emalpha}{\alpha} \cdot \past{\pi}(s,u) \\
    &\leq \frac{\emalpha}{\alpha} \cdot \frac{\rmax\cdot \omega}{\cur{m}} \cdot \sum_{s \in \past{V^*}} \past{d}(s) \sum_{u \in \past{V}}  \past{\pi}(s,u) \\
    &\quad + \frac{\emalpha}{\alpha} \cdot \frac{1}{\cur{m}} \cdot \sum_{s \in \past{V^*}}  \sum_{u \in \past{V}}  \past{\pi}(s,u) 
\end{flalign*}
\noindent where the second equality results from that $\past{d}(s)$ is positive and thus $\mceil{ \past{d}(s) \mathrm{\cdot} \rmax \mathrm{\cdot} \omega } \mathrm{\leq}  \past{d}(s) \mathrm{\cdot} \rmax \mathrm{\cdot} \omega \mathrm{+}1$. Due to the fact that $\sum_{u \in \past{V}} \past{\pi}(s,u) \mathrm{=} 1$, we have:
\begin{flalign*}
    \E\left[ \cur{c} \right] &\leq \frac{\emalpha}{\alpha} \cdot \frac{\rmax\cdot \omega}{\cur{m}} \cdot \sum_{s \in \past{V}} \past{d}(s) + \frac{\emalpha}{\alpha} \cdot \frac{\past{n}^*}{\cur{m}}\\
    &< \frac{\emalpha}{\alpha} \cdot (\rmax \cdot \omega + 1) = \bigO{\rmax \cdot \omega},
\end{flalign*}
where the second inequality comes from $\sum_{s \in \past{V}} \past{d}(s) \mathrm{=} \past{m} \mathrm{=} \cur{m}\mathrm{-}1$, and $\past{n}^* \leq \past{m}$ due to the definition of $\past{n}^*$.

Another part of cost is to expand $\cur{H}(\cur{u})$ by adding more random walks starting from $\cur{u}$ into $\cur{H}$ to satisfy the adequateness, we have:
\begin{flalign*}
  |\cur{H}(\cur{u})| - |\past{H}(\cur{u})| &= \mceil{\cur{d}(\cur{u}) \cdot \rmax \cdot \omega} - \mceil{\past{d}(\cur{u}) \cdot \rmax \cdot \omega} \\
  &= \bigO{\rmax \cdot \omega}
\end{flalign*}
\noindent
holds since $\cur{d}(\cur{u})\mathrm{=}\past{d}(\cur{u})\mathrm{+}1$. Finally, by setting $\rmax \mathrm{\cdot} \omega \mathrm{=} \Theta(1)$ according to SpeedPPR~\cite{WuGWZ21}, we achieve the result $\E[\cur{c}]\mathrm{=}O(1)$ and $|\cur{H}(\cur{u})|\mathrm{-}|\past{H}(\cur{u})|\mathrm{=}O(1)$, that is, we need to update only $O(1)$ random walks, and add only $O(1)$ new random walks.  

In Section \ref{sec:sample}, we presented an efficient sampling technique that takes only $O(1)$ to sample a random walk to update. Thus the cost of the procedure {\bf Sample} (Line \ref{algo:ins:upd}) is $O(1)$, due to $ \E[\cur{c}]\mathrm{=}O(1)$. When updating a random walk, we need to invoke procedure {\bf Walk-Restart} (Line \ref{algo:ins:upd-continue}). Since the expected length of a random walk with a decay factor $\alpha$ is $\bigO{1\mathrm{/}\alpha}\mathrm{=}O(1)$, the cost of {\bf Walk-Restart} is $O(1)$. Similarly, the cost of adding a new random walk is also $O(1)$.

From the above discussions, we know that {\bf Update-Insert} takes $O(1)$ expected time to update the index for each insertion. \done

\vspace{-2mm}
\header{\bf Proof of Theorem \ref{thm:delete-cost}.} As Algorithm~\ref{algo:del} shows, the cost of {\bf Update-Delete} also contains two parts. First, we need to trim $\past{H}$ to $\past{H'}$ by sampling random walks from $\past{H}(\cur{u})$ uniformly. Since the trimming process of $\cur{H}(\cur{u})$ in the deletion case is a reverse process of the expanding process of $\cur{H}(\cur{u})$ in the insertion case, we have the cost $|\past{H}(\cur{u})| \mathrm{-} |\cur{H}(\cur{u})|\mathrm{=}|\past{H}(\cur{u})| \mathrm{-} |\past{H'}(\cur{u})|\mathrm{=}O(\rmax\mathrm{\cdot}\omega)$.

Then, a deletion of edge $\cur{e}\mathrm{=}\tuple{u,v}$ leads to the rebooting of all random walks (except the trimmed random walks) that have records in $\past{C^E}(\cur{e})$. Let $\cur{c}$ denote the number of random walks that need to be repaired, the expectation of $\cur{c}$ is:
\begin{equation*}
\E\left[\cur{c}\right] \le \sum_{e \in \past{E}} \E\left[|\past{C^E}(e)|\right] \cdot \Pr\left[\cur{e} = e \right],
\end{equation*}
where the inequality is due to the fact that we only need to update a random walk once even if it passes through $\cur{e}$ multi-times and the records in $\past{C^E}$ that belong to trimmed random walks are also staled. Then, for each edge $e\mathrm{=}\tuple{u,v}\in \past{E}$, 
as a corollary of the definition of random walk, it follows that:
\begin{equation*}
\E\left[|\past{C^E}(e)|\right] = \frac{1}{\past{d}(\cur{u})} \cdot\E\left[|\past{C^V}(u)|\right].
\end{equation*}

Combining the equations above and Equation \ref{eqn:pr:del}, we obtain:
\begin{flalign*}
\E [\cur{c}] &\le \sum_{u \in \past{V}} \sum_{e \in \past{\Edge}(u)} \E\left[|\past{C^E}(e)|\right] \cdot \frac{1}{\past{m}}\\
&= \frac{1}{\past{m}} \cdot \sum_{u \in \past{V}} \sum_{e \in \past{\Edge}(u)}\frac{1}{\past{d}(\cur{u})} \cdot \E\left[|\past{C^V}(u)|\right] \\
&= \frac{1}{\past{m}} \cdot \sum_{u \in \past{V}} \E\left[|\past{C^V}(u)|\right].
\end{flalign*}

Given the above inequality, we can follow the same analysis steps as the insertion case, i.e., Equations \ref{eqn:relation-ppr}-\ref{eqn:relation-ppr-affected-edge} and the analysis after Equation \ref{eqn:relation-ppr-affected-edge} in the proof of Theorem~\ref{thm:insert-cost}, to derive that $\E[\cur{c}] \mathrm{=} O(\rmax\mathrm{\cdot}\omega)$.
As $\rmax\mathrm{\cdot}\omega \mathrm{=} \Theta(1)$ following SpeedPPR~\cite{WuGWZ21}, it yields that both $|\past{H}(\cur{u})| \mathrm{-} |\cur{H}(\cur{u})|\mathrm{=}O(1)$ and $\E [\cur{c}] \mathrm{=} O(1)$. Furthermore, as we described in the proof of Theorem~\ref{thm:insert-cost}, the cost of maintaining the index when updating or deleting a random walk can be bounded by $O(1)$. Hence, we get that {\bf Update-Delete} takes expected $O(1)$ time to update the index. \done

\section{Other Related Work}

\begin{table}[t]
\centering
\renewcommand{\arraystretch}{1.1}
\begin{tabular}{|l|l|r|r|r|}
    \hline
    {\bf Abbr.} & {\bf Name} & \multicolumn{1}{c|}{$\boldsymbol{n}$} & \multicolumn{1}{c|}{$\boldsymbol{m}$} & \multicolumn{1}{c|}{\bf Type}  \\ \hline
    {\em SF} & {\em Stanford}      & 281.9K    & 2.3M      & directed      \\ \hline
    {\em DB} & {\em DBLP}          & 317.1K    & 1.0M      & undirected    \\ \hline
    {\em YT} & {\em Youtube}       & 1.1M      & 3.0M      & undirected    \\ \hline
    {\em PK} & {\em Pokec}         & 1.6M      & 30.6M     & directed      \\ \hline
    {\em LJ} & {\em LiveJournal}   & 4.8M      & 69.0M     & directed      \\ \hline
    {\em OK} & {\em Orkut}         & 3.1M      & 117.2M    & undirected    \\ \hline
    {\em TW} & {\em Twitter}       & 41.7M     & 1.5B      & directed      \\ \hline
    {\em FS} & {\em Friendster}    & 65.6M     & 1.8B      & undirected    \\ \hline
\end{tabular}
\vspace{0mm}
\caption{Datasets. ($K=10^3, M=10^6, B=10^9$)}\label{tab:exp-data}
\end{table}

Personalized PageRank (PPR) was first proposed by Page et al.~\cite{Page1999}. They present a matrix-based definition of PPR and explore the Power-Iteration method. Given a source $s$, let $\vect{\pi}(s)$ denote the PPR vector of PPR scores of each node with respect to $s$. Let $\vect{A}$ be the adjacency matrix and $\vect{D}$ be a diagonal matrix where the $(u,u)$-th entry is the out-degree of $u$. Then, the following equation holds:
\begin{equation*}
\vect{\pi(s)} = \alpha \cdot \vect{e_s}+(1-\alpha)\cdot \vect{\pi(s)} \cdot \vect{D}^{-1}\vect{A},
    \end{equation*}
where $\vect{e_s}$ is a one-hot vector with only the $s$-th entry to be 1. The Power-Iteration method is still expensive and this motivates a series of works to solve the above linear system more efficiently~\cite{FujiNYSO12,MaeAIK14,ZhuFCY13,ShinJSK15,JungPSK17} via matrix related tricks, e.g., matrix decomposition. However, such methods are shown to be dominated by the Forward-Push + Monte-Carlo based methods as shown in \cite{WeiHX0SW18,WangWG0H20}.

Another line of solutions are local push methods. The Forward-Push~\cite{AnCL06} algorithm can be used to derive the answer of single-source PPR query. However, it provides no guarantee on the answers. The Backward-Push~\cite{JehW03,AnBCHMT07} is further proposed to derive the single-target PPR (STPPR) queries. Ohsaka et al.~\cite{OhsakaMK15} and Zhang et al.~\cite{ZhLG16} further design algorithms to update the stored Forward-Push results on dynamic graphs. In \cite{ZhLG16}, Zhang et al. further present algorithms to maintain the stored backward push results. Wang et al.~\cite{WangWG0H20} further present randomized Backward-Push to gain a better trade-off between query efficiency and accuracy. However, all these methods cannot be applied to answer ASSPPR/ASSPPR top-$k$ queries. 

\begin{figure*}[t!]
\centering
\vspace{-2mm}
\begin{small}
\begin{tabular}{cccc}
\multicolumn{4}{c}{
\hspace{-6mm}\includegraphics[height=2.5mm]{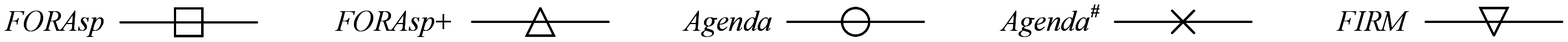}}
\\[-1mm]
  \hspace{-2mm} \includegraphics[height=27mm]{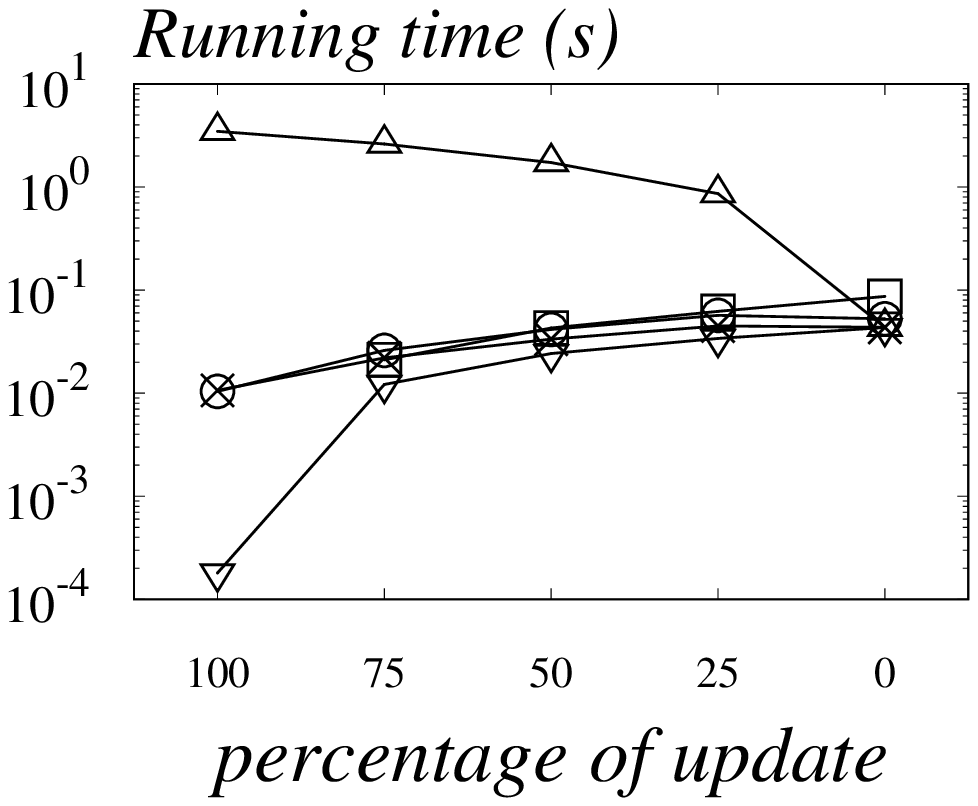} &
  \hspace{-2mm} \includegraphics[height=27mm]{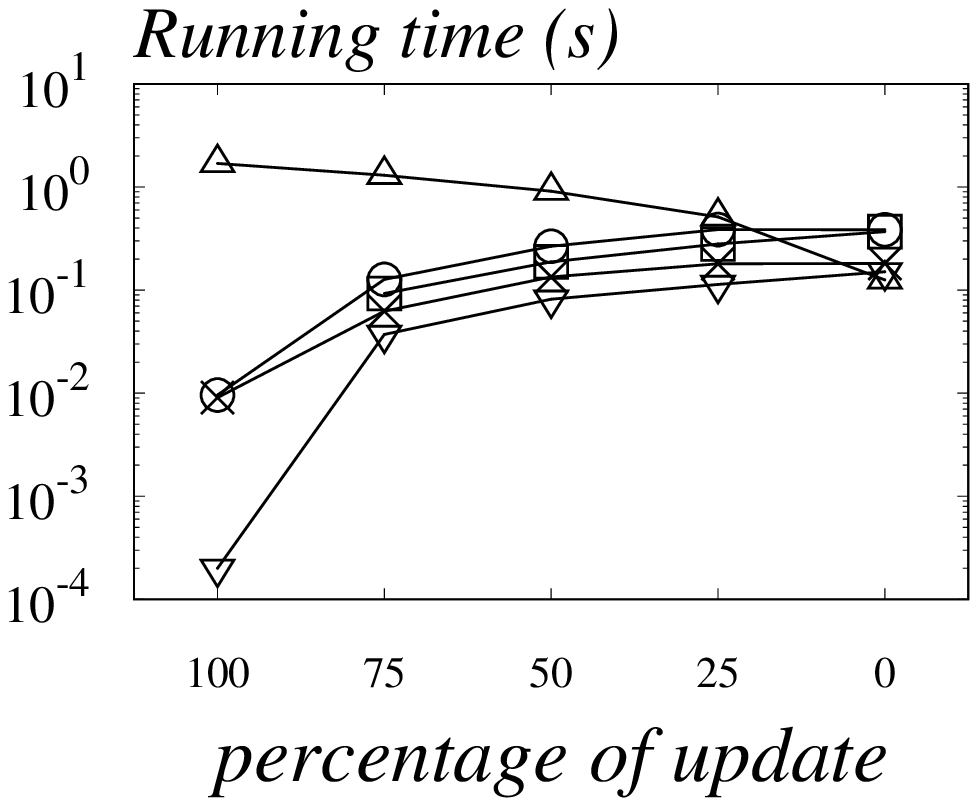} &
  \hspace{-2mm} \includegraphics[height=27mm]{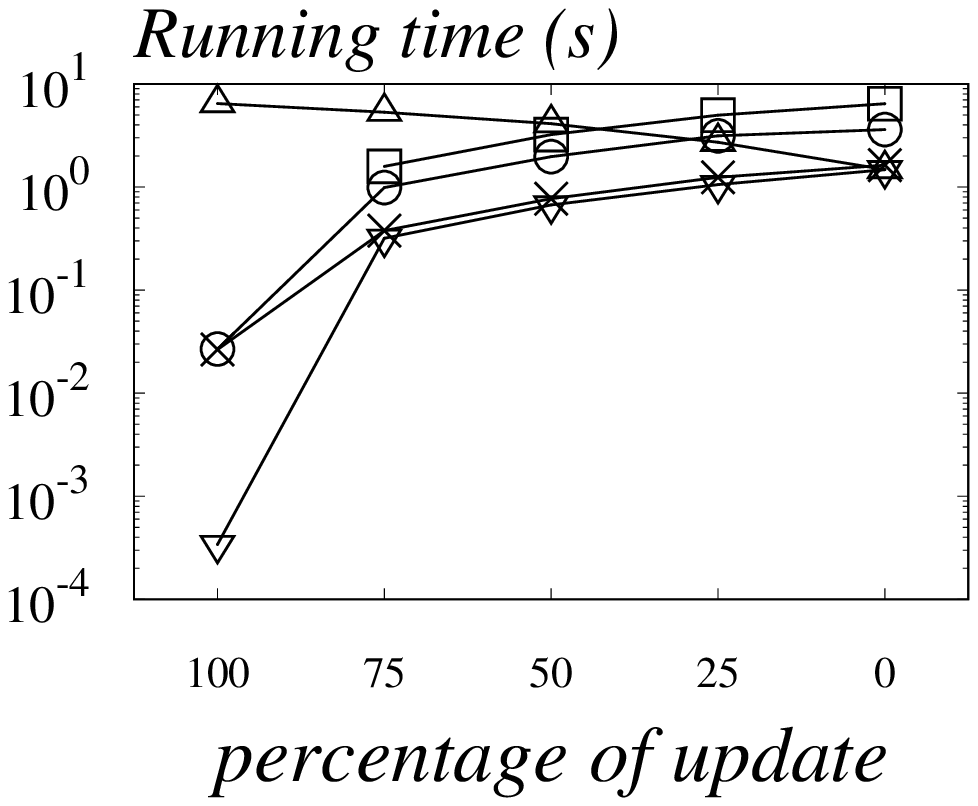} &
  \hspace{-2mm} \includegraphics[height=27mm]{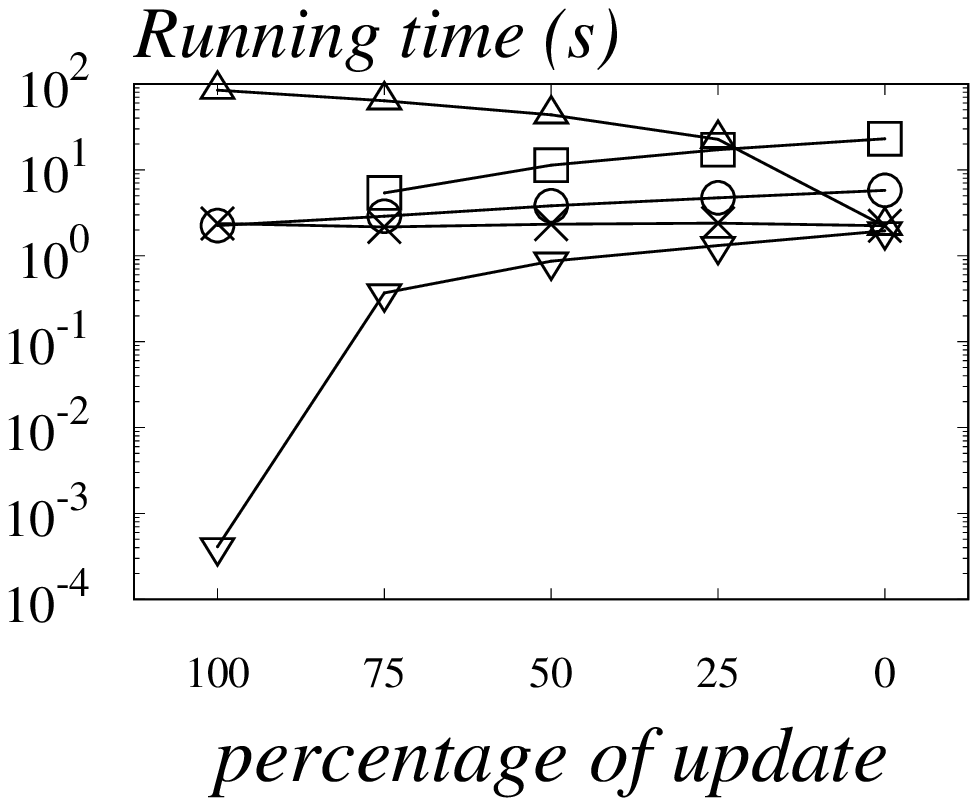}
\\[-2mm]
  \hspace{-4mm} (a) Stanford &
  \hspace{-4mm} (b) DBLP &
  \hspace{-4mm} (c) Youtube &
  \hspace{-4mm} (d) Pokec 
\\[-1mm]
  \hspace{-2mm} \includegraphics[height=27mm]{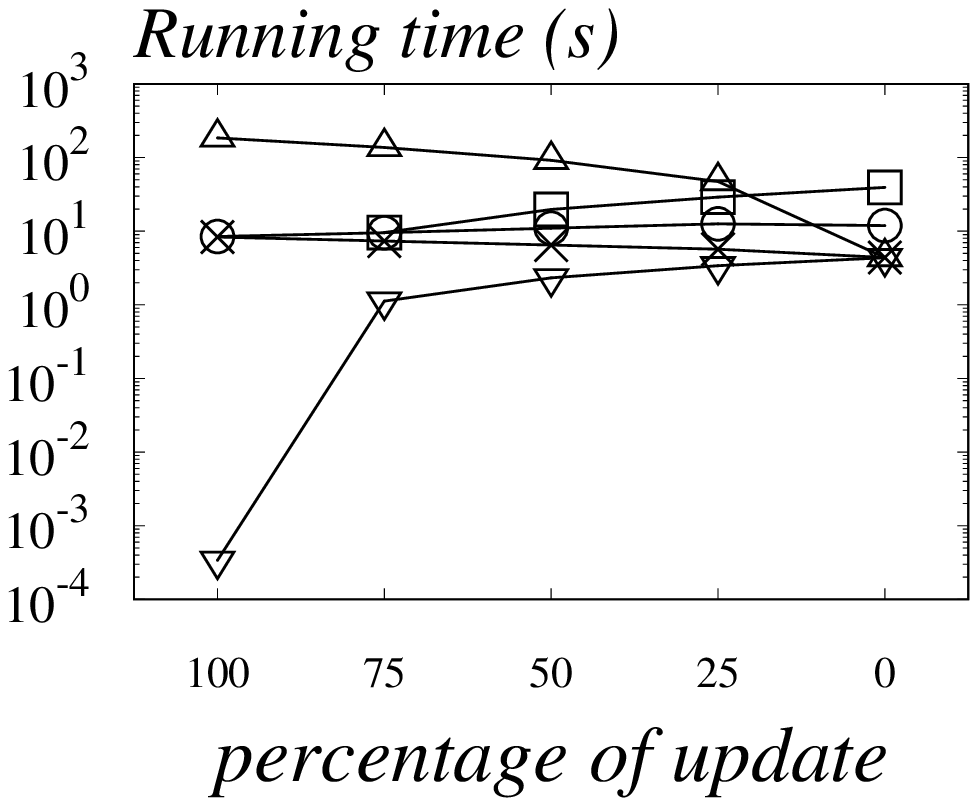} &
  \hspace{-2mm} \includegraphics[height=27mm]{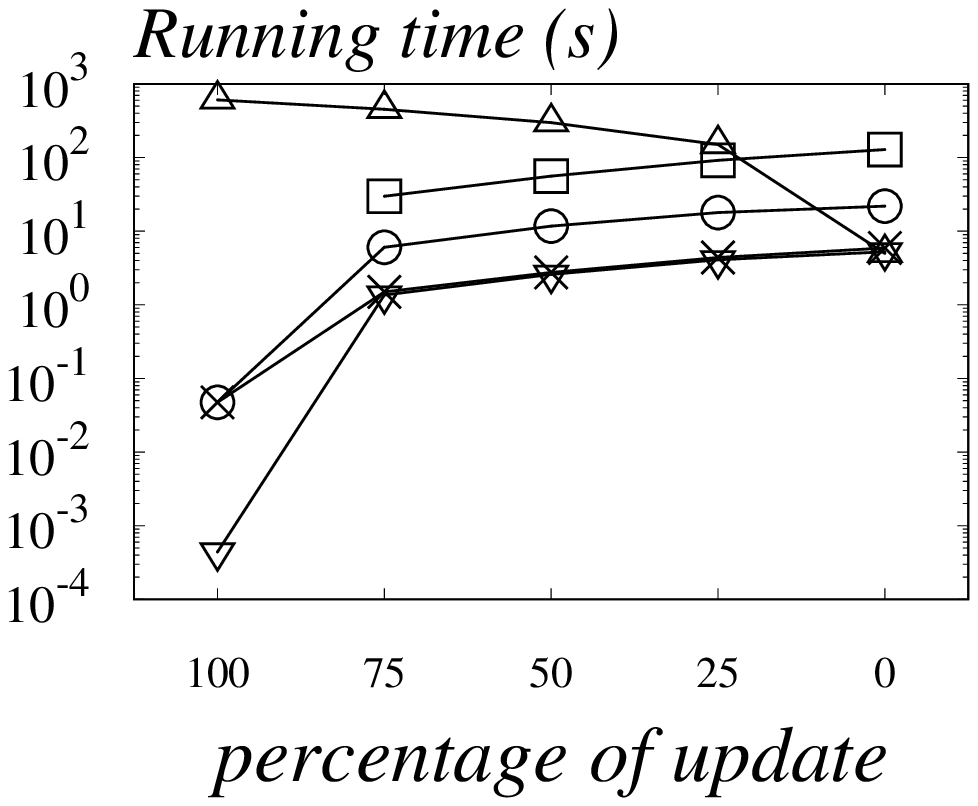}&
  \hspace{-2mm} \includegraphics[height=27mm]{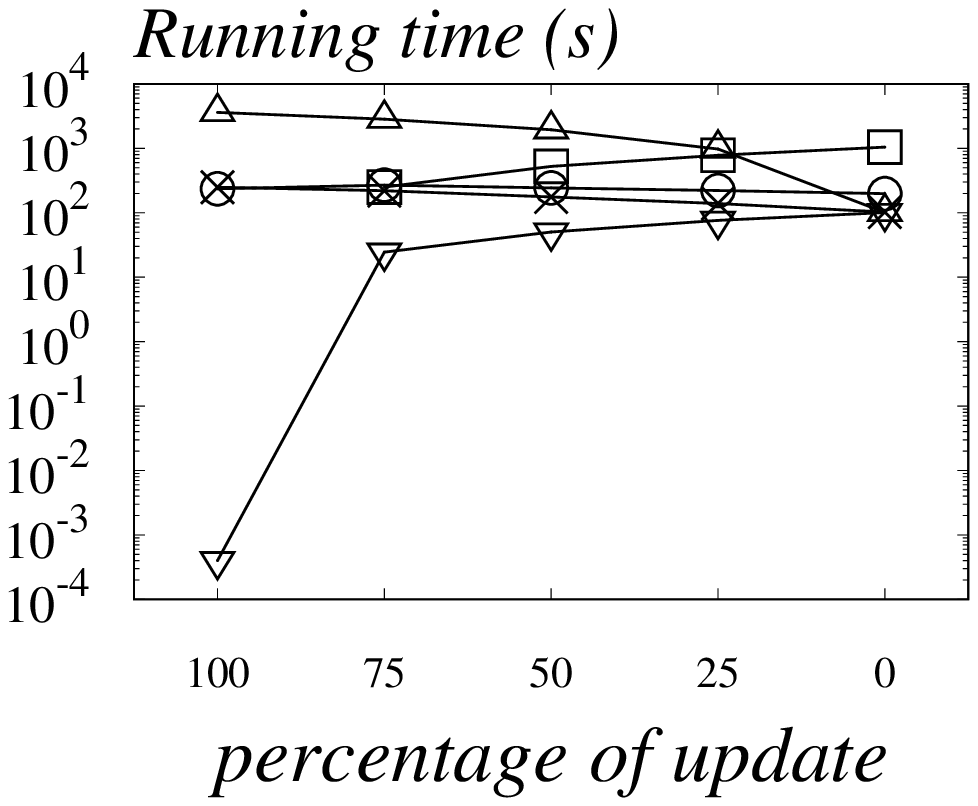} &
  \hspace{-2mm} \includegraphics[height=27mm]{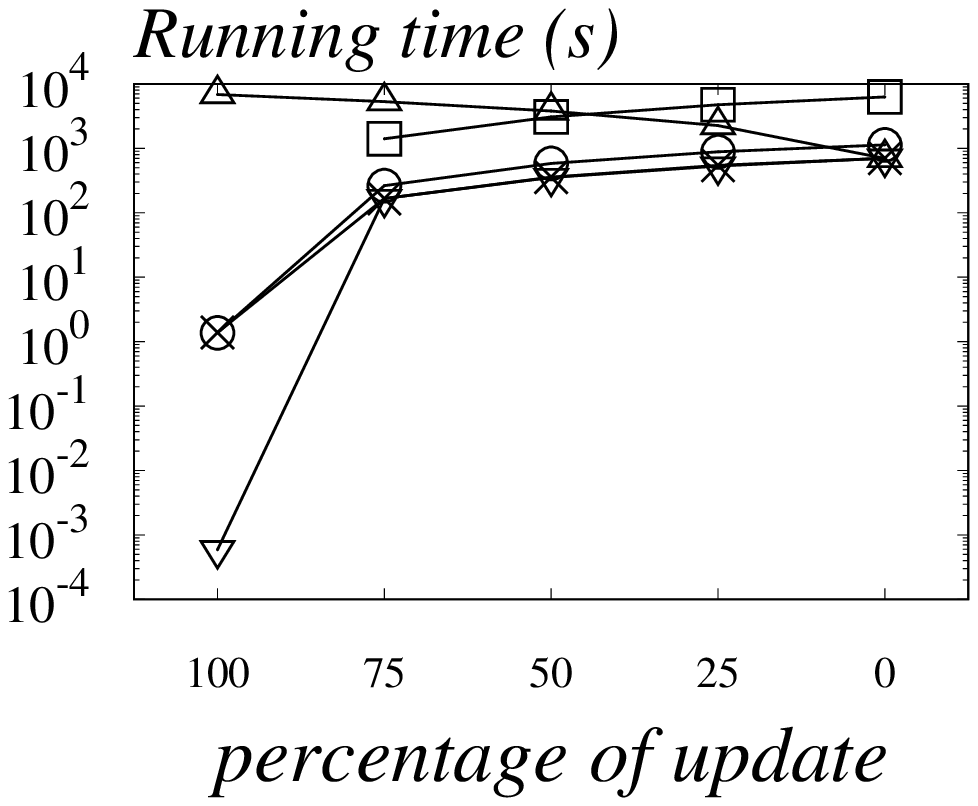}
\\[-2mm]
  \hspace{-4mm} (e) LiveJournal &
  \hspace{-4mm} (f) Orkut &
  \hspace{-4mm} (g) Twitter &
  \hspace{-4mm} (h) Friendster
\end{tabular}
\caption{Average processing time with ASSPPR queries}  \label{fig:perf:full}
\end{small}
\end{figure*}

\begin{figure*}[t]
\centering
\vspace{-4mm}
\begin{small}
\begin{tabular}{cccc}
  \hspace{-2mm} \includegraphics[height=27mm]{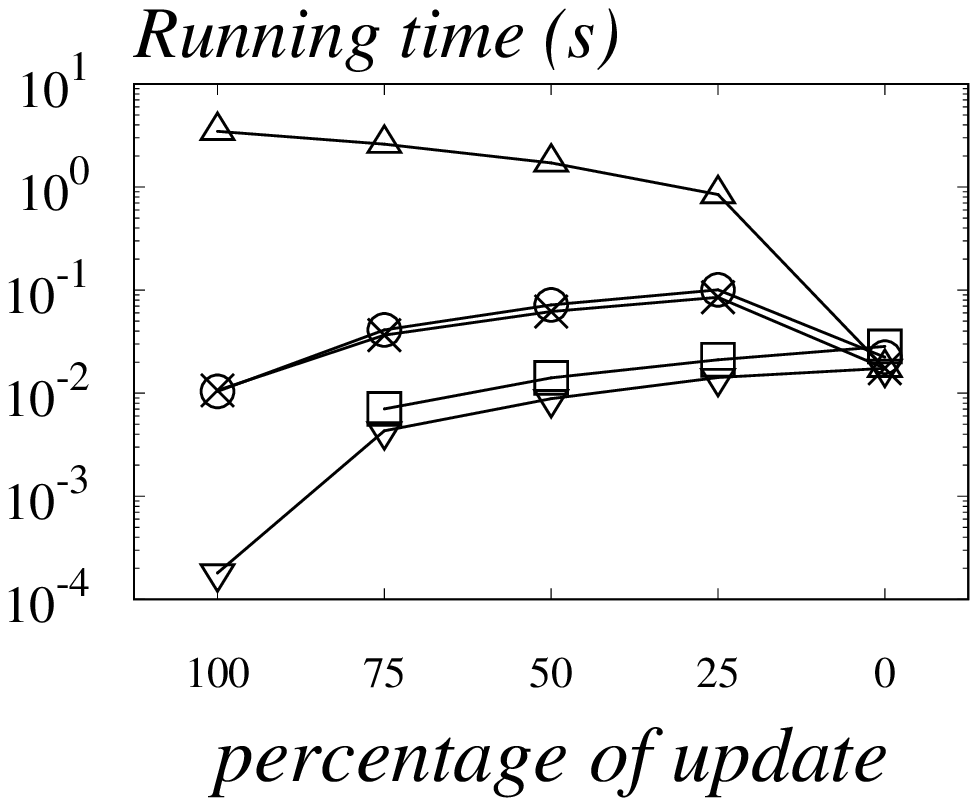} &
  \hspace{-2mm} \includegraphics[height=27mm]{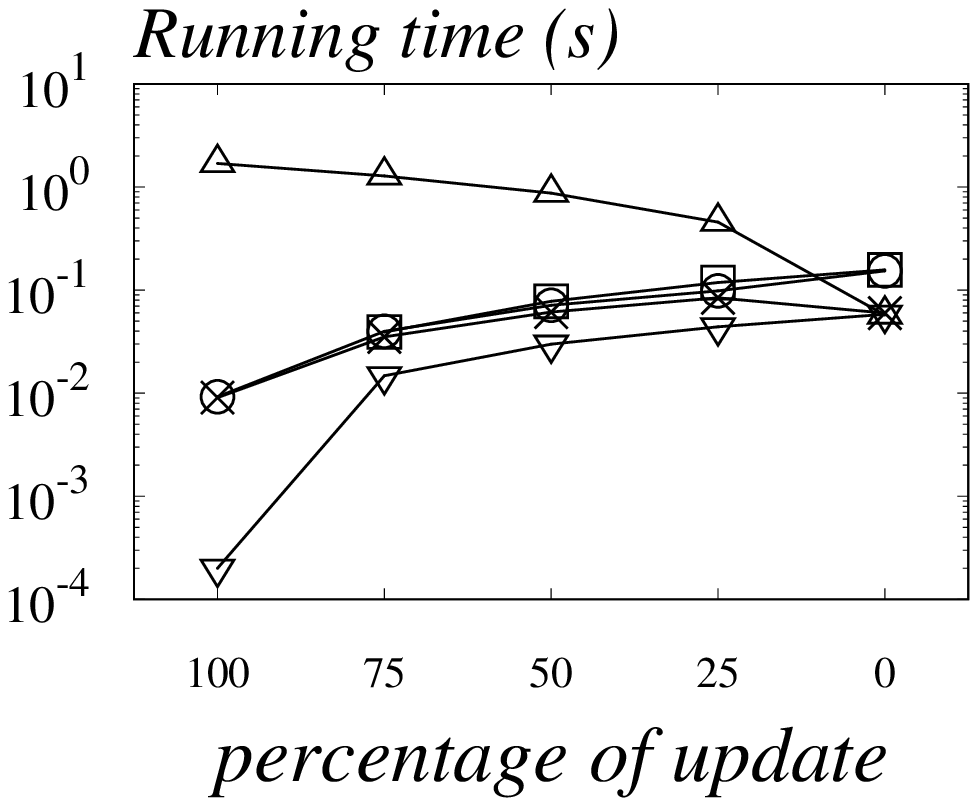}&
  \hspace{-2mm} \includegraphics[height=27mm]{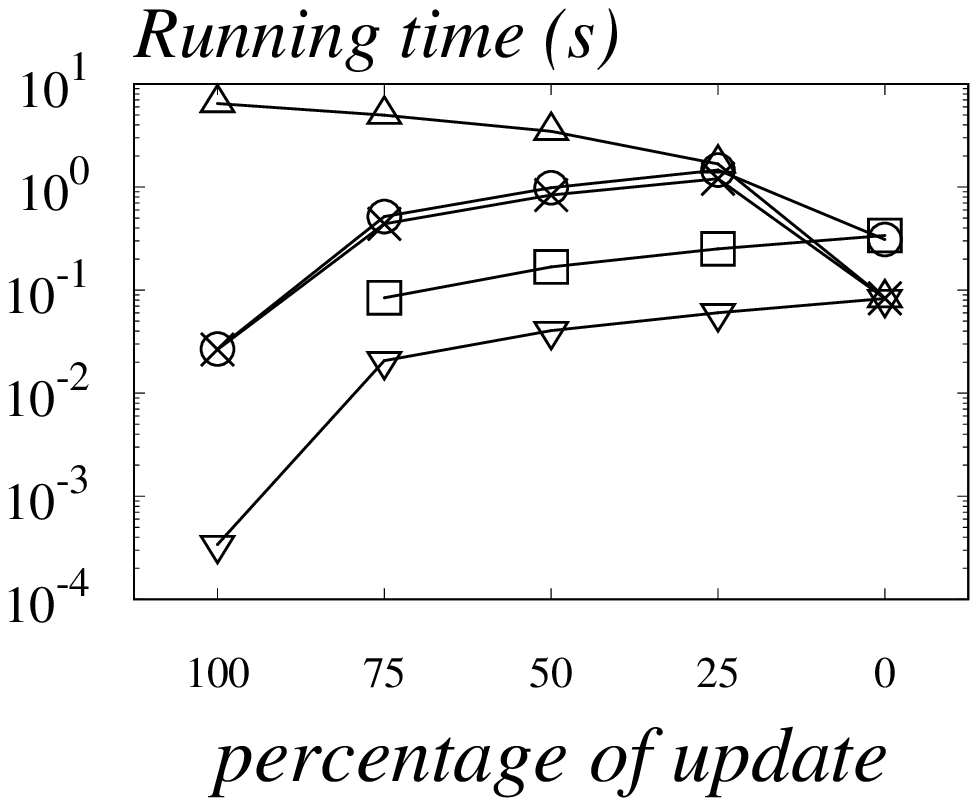} &
  \hspace{-2mm} \includegraphics[height=27mm]{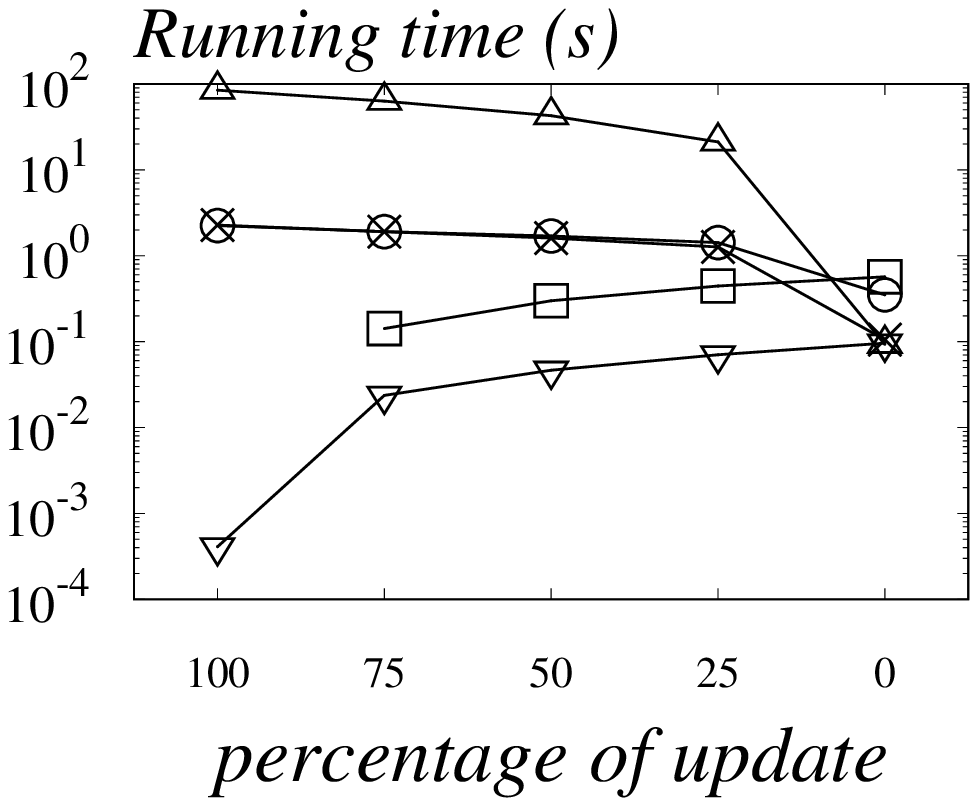}
\\[-2mm]
  \hspace{-4mm} (a) Stanford &
  \hspace{-4mm} (b) DBLP &
  \hspace{-4mm} (c) Youtube &
  \hspace{-4mm} (d) Pokec 
\\[-1mm]
  \hspace{-2mm} \includegraphics[height=27mm]{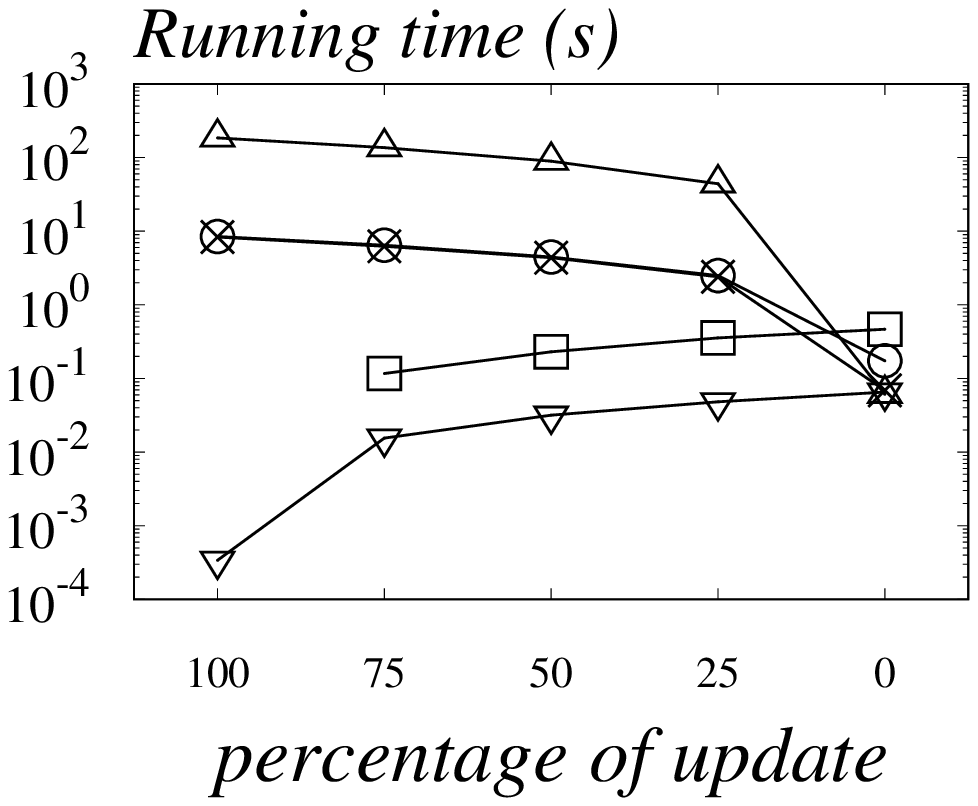} &
  \hspace{-2mm} \includegraphics[height=27mm]{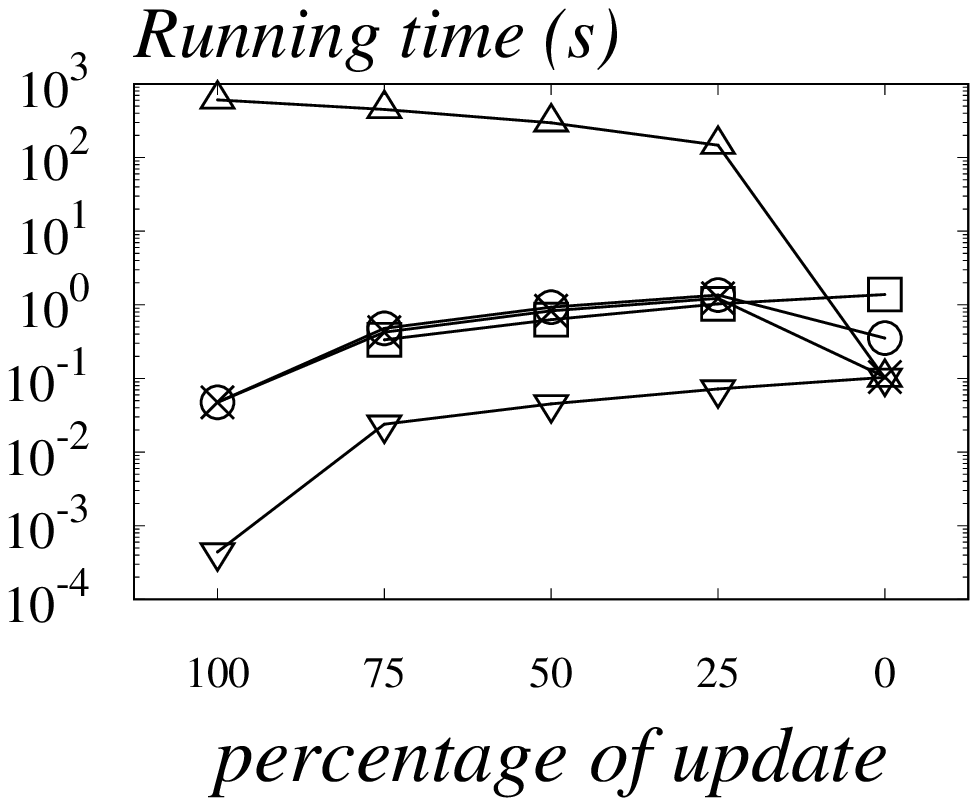}&
  \hspace{-2mm} \includegraphics[height=27mm]{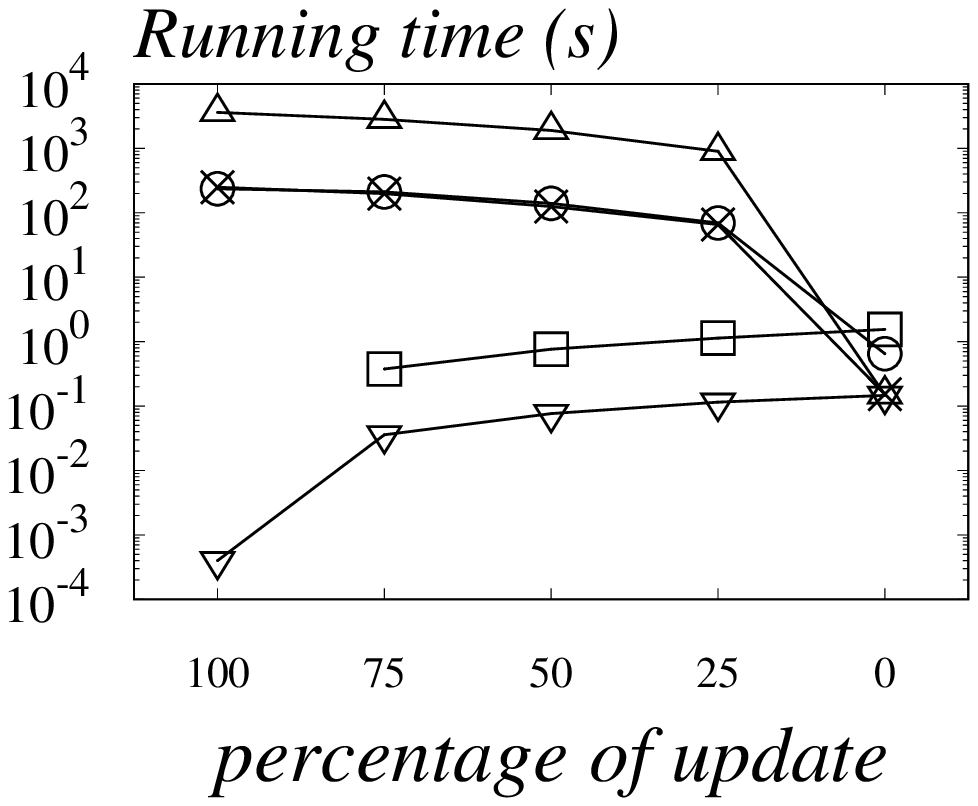} &
  \hspace{-2mm} \includegraphics[height=27mm]{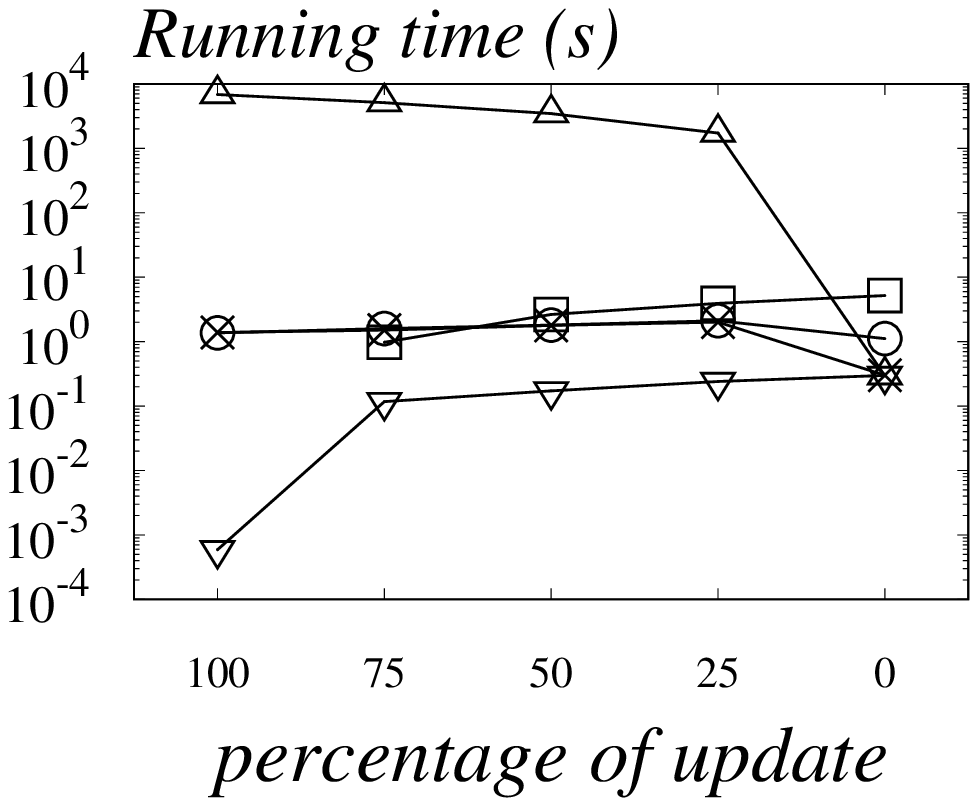}
\\[-2mm]
  \hspace{-4mm} (e) LiveJournal &
  \hspace{-4mm} (f) Orkut &
  \hspace{-4mm} (g) Twitter &
  \hspace{-4mm} (h) Friendster
\end{tabular}
\caption{Average processing time with ASSPPR top-$k$ queries.} \label{fig:perf:topk}
\end{small}
\end{figure*}

To gain an approximation guarantee, Monte-Carlo methods~\cite{FogRCS05} are proposed to derive approximate estimations. However, Monte-Carlo methods alone are still too slow, which motivates existing solutions to combine local push algorithms and Monte-Carlo methods to gain better query efficiency while still providing approximation guarantee. In particular, Lofgren et al.~\cite{LofBGS14,LofBG15} and Wang et al.~\cite{WangTXYL16,WangT18} present solutions to combine random walk and Backward-Push to improve the query performance of pairwise PPR queries with approximation guarantees. Later, FORA~\cite{WYWXWLYT19},  ResAcc~\cite{LinWXW20}, and SpeedPPR~\cite{WuGWZ21} are further proposed to combine the Forward-Push and random walks to improve the query performance for ASSPPR queries. The index-based version of FORA and SpeedPPR, dubbed as FORA+ and SpeedPPR+, respectively, are shown to incur high update costs as in our experiment. Besides, ResAcc is an index-free method while its query processing is not as fast as such index-based solutions.

There also exist a line of research works, e.g., \cite{FujiNYSO12,FujiNSMO13,WeiHX0SW18,WYWXWLYT19} on efficient top-$k$ PPR query processing. The state-of-the-art approach is FORA+, which achieves the best query efficiency as shown in~\cite{WYWXWLYT19}, but it is an index-based method. Previously there exist no efficient algorithms to support dynamic index update and our {\oursolution} fills this gap. The state-of-the-art index-free method is TopPPR~\cite{WeiHX0SW18}, which combines Forward-Push, random walk, and Backward-Push to answer top-$k$ PPR queries with precision guarantees. However, TopPPR is only designed for top-$k$ queries and cannot support SSPPR queries. In contrast, with the same random walk index, our {\oursolution} supports both efficient ASSPPR and ASSPPR top-$k$ queries. 

Finally, there exist research works on parallelizing PPR computations with multi-core~\cite{WangWZ19}, GPU~\cite{GuoLST17,ShiYJXY19}, or in distributed environment~\cite{GuoCCLL17,SarmaMPU13,Luo19,LinWXW20,HouCWW21}. These works are orthogonal to ours.

\section{Experiments}
Next, we experimentally evaluate our {\oursolution} against alternatives. All experiments are conducted on an AWS x1.16xlarge cloud server with 64vCPUs clocked at $2.3$GHz and $976$GB memory. All codes are implemented in C++ and compiled with full optimization.

\subsection{Experimental Settings}\label{sec:setting}
We compare our method against four solutions.
FORAsp is the method whose number of random walks is set to $O(m)$ following SpeedPPR~\cite{WuGWZ21} whereas its workflow is identical to the original FORA~\cite{WYWXWLYT19} because Power-Push of SpeedPPR on evolving graphs is not as efficient as that on static graphs. FORAsp+ is the index-based version of FORAsp. The state-of-the-art solution for evolving graphs, Agenda~\cite{MoLuo21}, is also included. We further include {\agendastar}, a variant of Agenda, which has been discussed in Section \ref{sec:sol:dyn}.

\header
{\bf Datasets and Metrics.} We use 8 benchmark datasets that can be obtained from public sources SNAP\cite{snapnets} and Konect\cite{konectdata} and are frequently used in previous research works on PPR, e.g., \cite{WYWXWLYT19,LofBG15,WuGWZ21,MoLuo21}, as shown in Table~\ref{tab:exp-data}. To measure the performance of the solutions on evolving graphs, for each dataset, we randomly shuffle the order of edges and divide it into two parts. The first part which has $90\%$ edges ($50\%$ edges for Twitter and Friendster to reduce the running time on cloud servers) will be used to build the initial graph. Then, we generate workloads each consisting of 100 updates/queries. An update will be either {\em (i)} an insertion of an edge selected randomly from the rest part of the edges, or {\em(ii)} a deletion of an edge selected randomly from the initial graph. A workload with update percentage $x\%$ means that it contains $x$ updates and $(100\mathrm{-}x)$ queries.

\header
{\bf Parameter Settings.} Following previous work~\cite{WYWXWLYT19,WuGWZ21}, we set $\alpha\mathrm{=}0.2$, $\epsilon\mathrm{=}0.5$, $\delta\mathrm{=}1\mathrm{/}n$ and $p_f\mathrm{=}1\mathrm{/}n$ by default. In addition, we
set Agenda according to~\cite{MoLuo21} such that $\theta\mathrm{=}0.5$ and $\rmax^b\mathrm{=}\cur{d}(\cur{u})\mathrm{/}\cur{m}$ on undirected graphs or $\rmax^b\mathrm{=}1\mathrm{/}\cur{n}$ on directed graphs. For top-k queries, we set $k\mathrm{=}500$. To speed up query processing of index-based FORA-like methods, we balance the time cost between Forward-Push and refining phases by setting $\rmax\mathrm{\cdot}\omega \mathrm{=} \beta\mathrm{/}\alpha$, where $\beta$ is a parameter depending on the dataset.

\subsection{Performance of {\oursolution}}
{\bf General Performance.} We first reveal the general performance of {\oursolution}, Figure~\ref{fig:perf:full} shows the performance under different workloads consisting of edge updates and ASSPPR queries. In a word, our solution {\oursolution} outperforms all alternatives under an arbitrarily mixed workload for all the datasets. The advantage of {\oursolution} is more prominent for workloads that consist of edge updates and ASSPPR top-$k$ queries as shown in Figure~\ref{fig:perf:topk}. This is because the query cost of a top-$k$ query is much slighter than a full ASSPPR query so the update cost will significantly affect the performance. Therefore, {\oursolution} has great practical value since ASSPPR top-k query is widely used for web-search, recommendation systems, and other scenarios. Besides, the essence of an algorithm for top-k queries is to provide a rough but adequate precision of SSPPR. Thus, {\oursolution} can also be applied to the scenarios which need a loose $(\epsilon,\delta)$-approximation guarantee (e.g. $\delta \mathrm{=} O(1)$) on evolving graphs to improve their performance.

\begin{figure}[t]
    \centering
    \includegraphics[height=2.2mm]{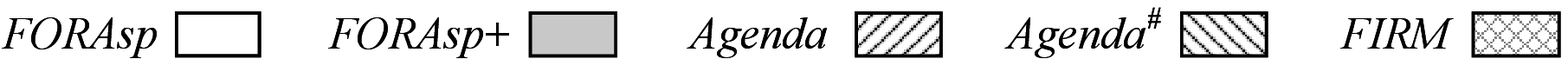}\\[0mm]
    \includegraphics[height=2.8cm]{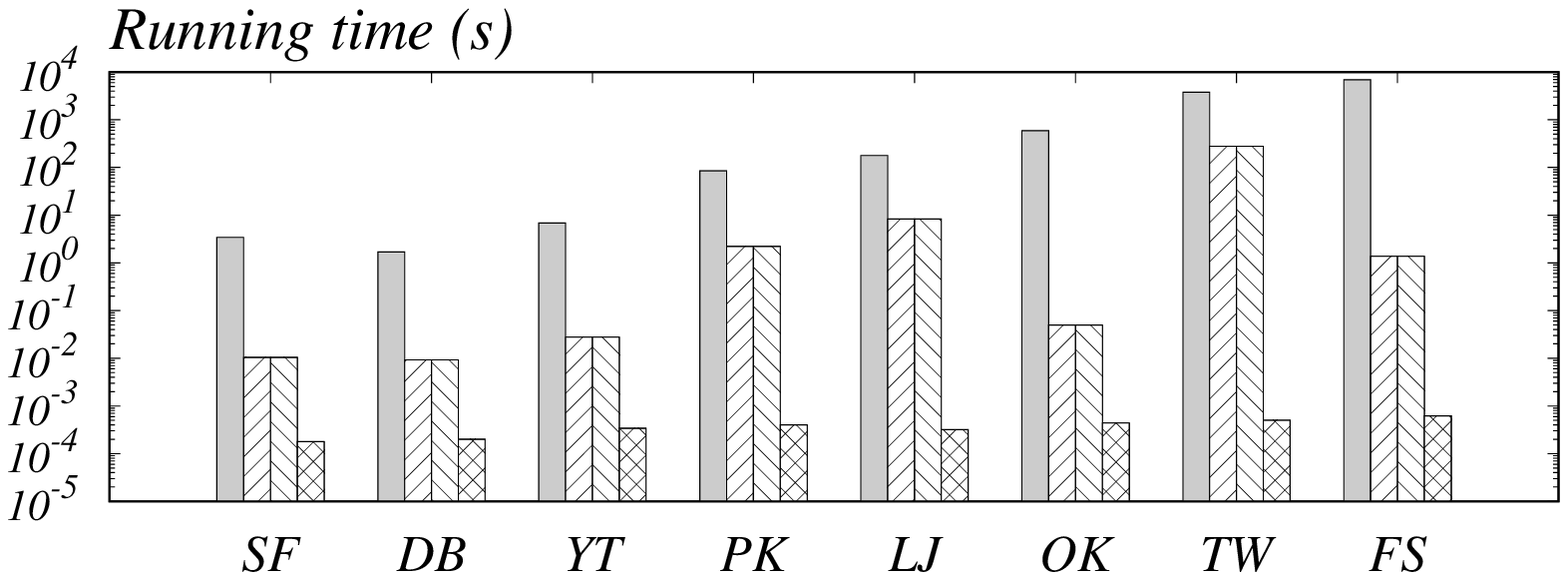}
    \caption{Average Update Time}\label{fig:perf:upd}
\end{figure}

\begin{figure}[t]
    \centering
    \includegraphics[height=2.8cm]{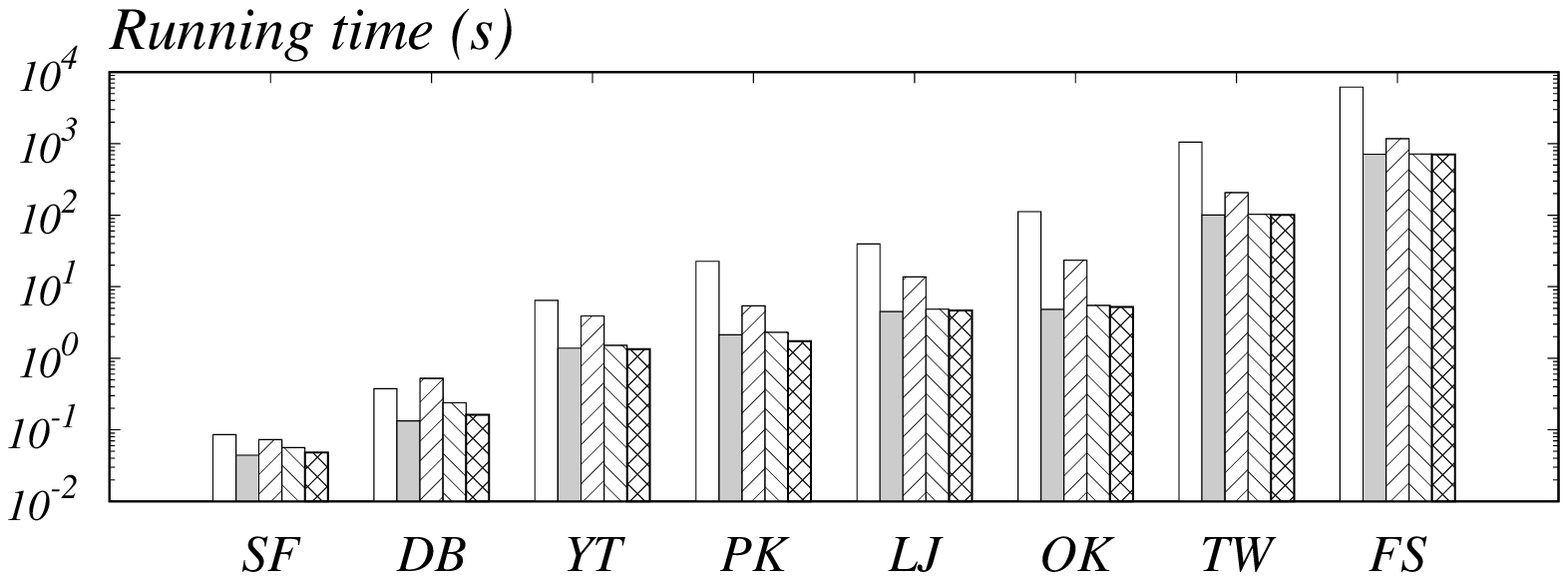}
    \caption{Average Query Time of ASSPPR Queries} \label{fig:perf:qfull}
\end{figure}

\begin{figure}[t]
    \centering
    \includegraphics[height=2.8cm]{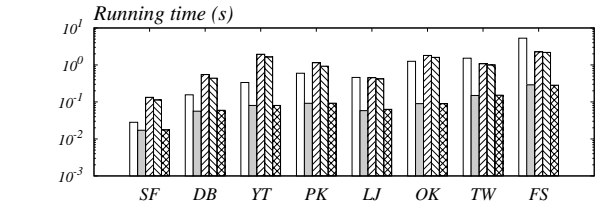}
    \caption{Average Query Time of ASSPPR Top-k Queries} \label{fig:perf:qtopk}
    \vspace{2mm}
\end{figure}

\header
{\bf Update Performance.} We evaluate the efficiency of {\oursolution} against index-based alternatives for updating the index structure. Figure~\ref{fig:perf:upd} shows the average processing time for each update under a workload with $50\%$ update. Observe that FORAsp+ has the worst update performance since it simply rebuilds its index which leads to a prohibitive computation. Compared to FORAsp+, Agenda does improve the update performance. Note that, the update process of Agenda is just to trace the inaccuracy of its index, the affected random walks will be reconstructed during query processing. {\agendastar} has the same update process as Agenda. Our solution, {\oursolution}, is orders of magnitude faster than FORAsp+ and Agenda. Moreover, {\oursolution} has a similar time consumption among all datasets which confirms that our solution takes $O(1)$ time to maintain its index for each update. In contrast, the update time of FORAsp+ and Agenda/{\agendastar} increases notably with graphs becoming larger.

\header
{\bf Full Query Performance.} We compare the average time for ASSPPR query processing under a workload with an update percentage of $50$ to reflect the additional cost of the lazy-update strategy of Agenda/{\agendastar}. Figure~\ref{fig:perf:qfull} shows the average query time of ASSPPR queries. Not surprisingly, FORAsp is the slowest one to answer ASSPPR queries because it is an index-free approach. {\oursolution} is as fast as FORAsp+ for query processing and achieves over 10x speed-up over FORAsp on most datasets. Agenda is faster than FORAsp but slower than FORAsp+ and {\oursolution} because it may have to reconstruct some random walks before query processing, and need more computation in Forward-Push phase to provide the same approximation guarantee since it admits inaccuracy tolerance when updating the index. {\agendastar} takes slightly more running time than FORAsp+ and {\oursolution}, as it needs to apply the lazy-update scheme during query processing. However, with the avoidance of additional computation in Forward-Push phase, it is surely faster than Agenda.

\header
{\bf Top-$\boldsymbol{k}$ Query Performance.} The situation is not quite the same when processing ASSPPR top-$k$ queries. As Figure~\ref{fig:perf:qtopk} shows, Agenda is even slower than FORAsp on some datasets. To explain, the process to answer top-$k$ queries in~\cite{WYWXWLYT19} repeats invoking Forward-Push with a rough $\rmax'\mathrm{>}\rmax$, refining the temporary result to provide a $(\epsilon, \delta')$-approximation guarantee where $\delta'\mathrm{>}\delta$ is a rough threshold and checking whether $\delta'$ is enough to bound the top-$k$ PPR scores. After each time Forward-Push is invoked, Agenda must check and fix the inaccuracy of its index, making the performance analysis in~\cite{MoLuo21} not applicable anymore. 

{\agendastar} suffers from the same problem as well. 
Besides, it has to be mentioned that in Figure~\ref{fig:perf:topk} the query performance of {\agendastar} (with $100\%$ query) is almost as fast as FORAsp+ and {\oursolution}, while in Figure~\ref{fig:perf:qtopk} its performance is much worse. It seems paradoxical at first glance. To explain, it is a tricky optimization for a workload with a low update rate. We can easily maintain an upper bound for the total error of the current index, and then if the upper bound is below error tolerance, we are allowed to skip the lazy-update phase. Thus, the lazy-update process is never invoked under a pure query workload. However, under mixing workloads, the lazy-update process is frequently invoked, thus becoming the bottleneck of {\agendastar}. In contrast, our solution still keeps the same query performance as FORAsp+ for top-$k$ queries and achieves over 10x speed-up over FORAsp.

\begin{figure}[t]
    \centering
    \includegraphics[height=2.2mm]{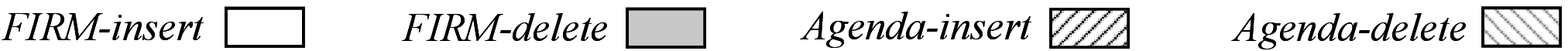}
    \vspace{0mm}
    \includegraphics[height=2.8cm]{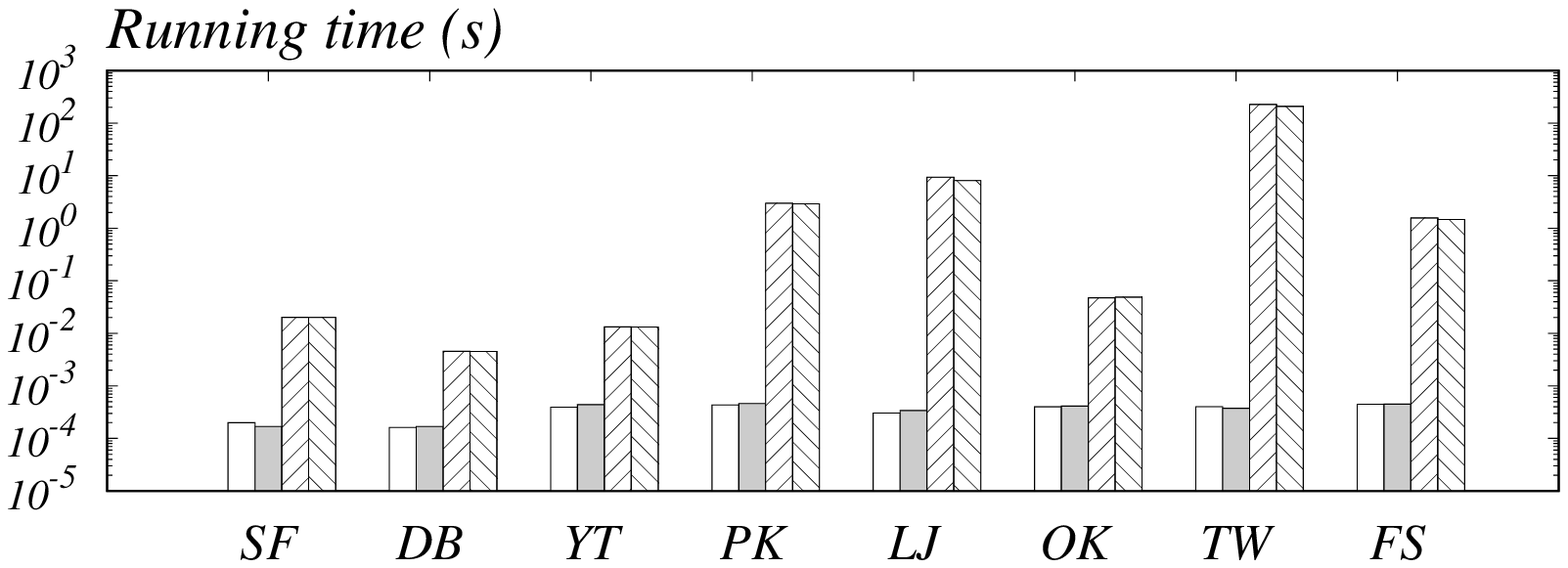}
    \vspace{-1mm}
    \caption{Average cost of insertion and deletion} \label{fig:perf:mixing}
    \vspace{1mm}
\end{figure}

\header
{\bf Performance of Insertion and Deletion.} To compare the updating cost of edge insertion with that of edge deletion, we respectively run 1000 insertions and 1000 deletions on each dataset, by using {\oursolution} and Agenda. We omit FORAsp+ because its update scheme is the same for edge insertion and deletion. {\agendastar} is also omitted because it takes the same updating cost as Agenda as shown in Figure \ref{fig:perf:upd}. Figure \ref{fig:perf:mixing} shows the average running time of edge insertion and deletion. As we can observe, with our {\oursolution}, the update cost for edge insertion is almost equal to that for edge deletion, which confirms our theoretical analysis that both of them have $O(1)$ cost. Agenda also has a similar trend, consistent with the analysis in~\cite{MoLuo21}.

\begin{figure}[!t]
	\centering
	\vspace{-2mm}
	\begin{small}
		\begin{tabular}{cc}
        \multicolumn{2}{c}{\hspace{-1mm}\includegraphics[height=2.2mm]{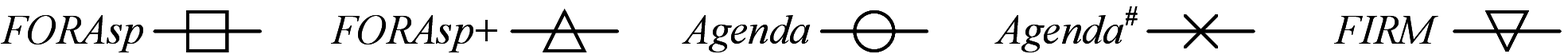}}  \\[0mm]
			\hspace{-2mm} \includegraphics[height=27mm]{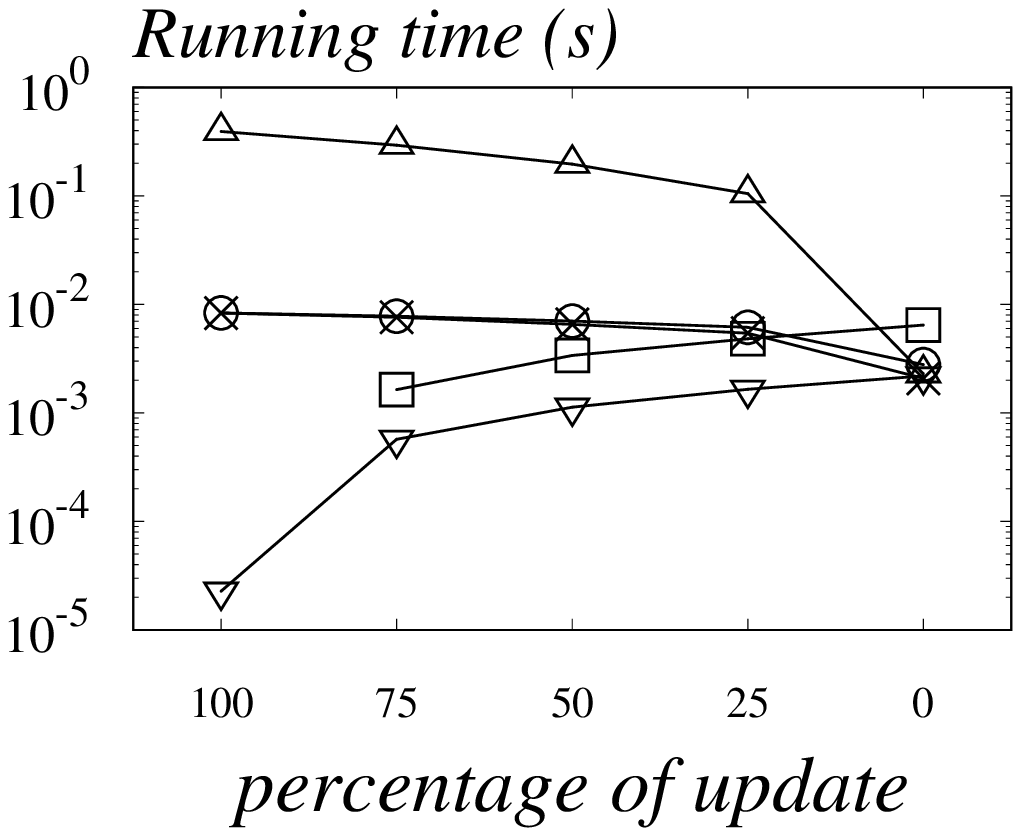} &
			\hspace{-2mm} \includegraphics[height=27mm]{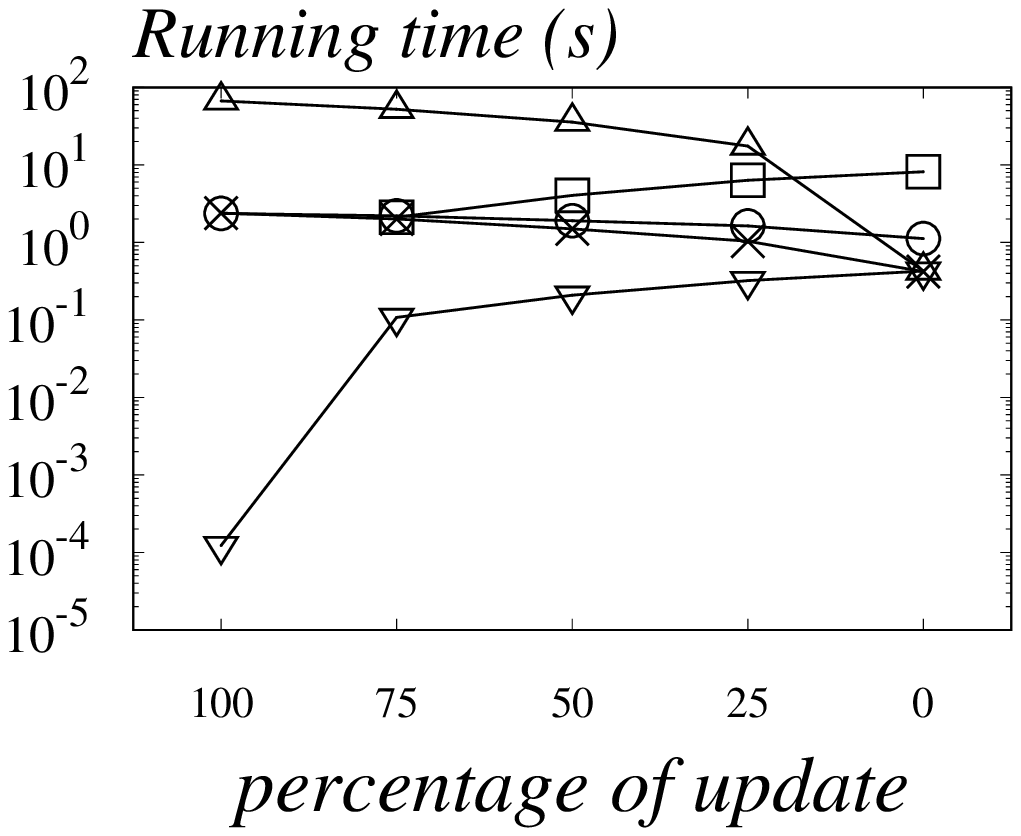} \\[-1.5mm]
			
			\hspace{-1.5mm} (a) Digg  &
			\hspace{-1.5mm} (b) Flickr \\
			
			\hspace{-2mm} \includegraphics[height=27mm]{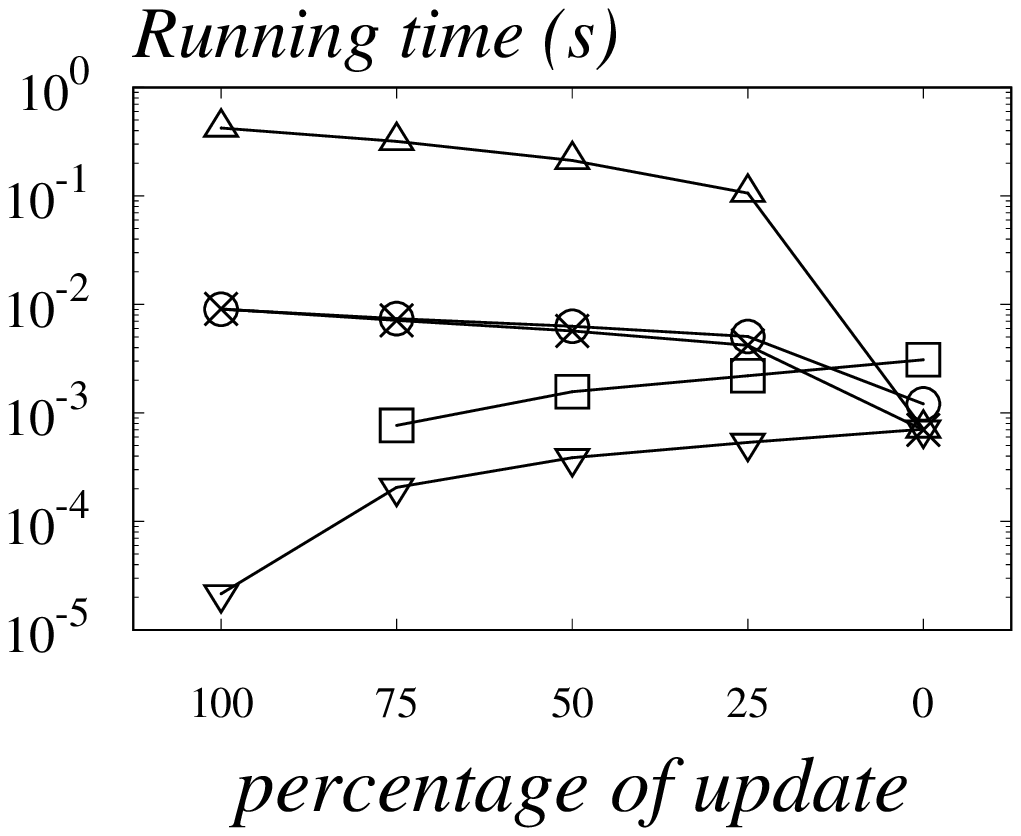} &
			\hspace{-2mm} \includegraphics[height=27mm]{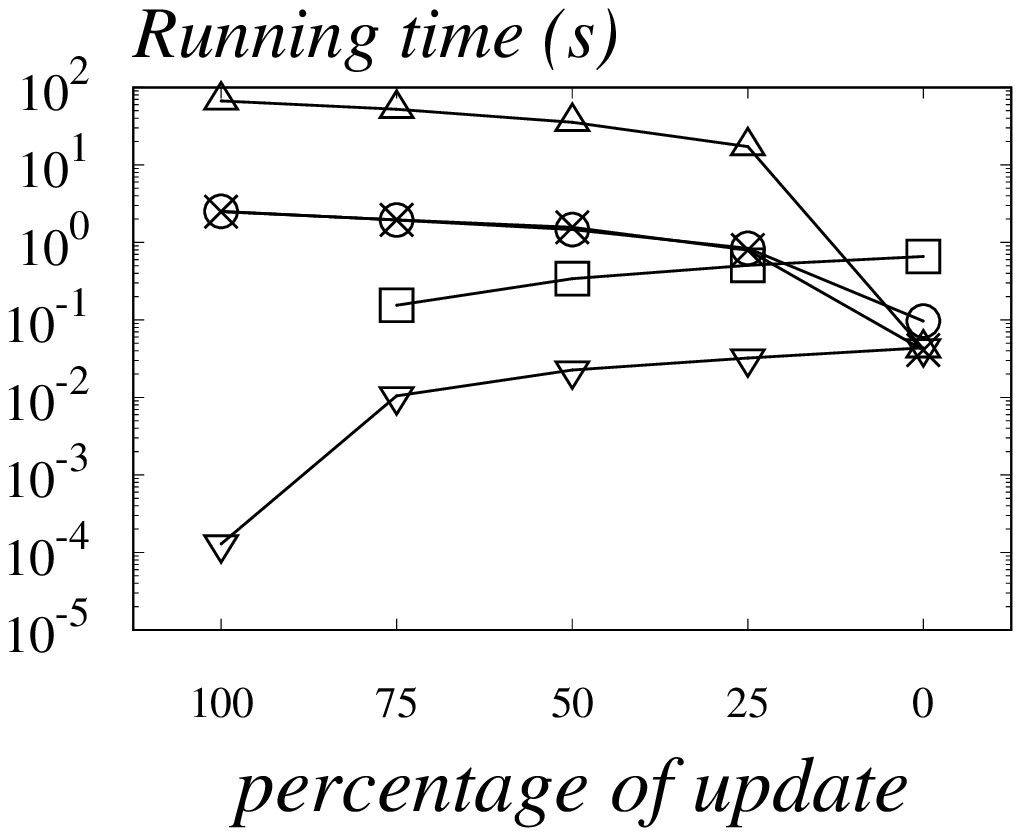}
			\\[-1.5mm]
			\hspace{-1mm} (c) Digg Top-k &
			\hspace{-1mm} (d) Flickr Top-k
		\end{tabular}
		\caption{Average processing time on temporal graphs.}  \label{fig:timegraph}
	\end{small}
	\vspace{0mm}
\end{figure}

\header
{\bf Performance on Real-World Temporal Graphs.} To examine the effectiveness of FIRM in real-world scenarios, we have run FIRM and alternatives on two temporal social networks, Digg (0.28M nodes and 1.7M edges) and Flickr (2.3M nodes and 33.1M edges), where edge timestamps (i.e., when the edge is built) are provided. The two graphs can also be downloaded at \cite{konectdata}. To simulate the evolving process, we design the experiment as follows: We first sort all edges by the edge timestamp in ascending order and take the first $90\%$ edges as the initial graph; then remaining edges will be sequentially added to the graph according to their timestamps. Figure \ref{fig:timegraph} shows the average processing time of our {\oursolution} and its competitors for the ASSPPR queries. From the figures, we can see that {\oursolution} still keeps superb efficiency on real-world evolving graphs with orders of magnitude speedup. We further examine the update cost of {\oursolution} and Agenda on these two graphs which are generated under the random arrival model rather than sorted by their actual timestamps. \crver{The results are in our technical report \cite{firm-tr}, which shows that {\oursolution} (as well as Agenda) has similar update performance under these two edge arriving settings, with negligible gap.} \trver{The results are shown in Table \ref{tab:timegraph_cmp}. We can observe that the updating cost of FIRM between real-world timestamp and random arrival model has a gap around $25\%$ which is not significant. Besides, Agenda has a similar trend for the update cost under these two edge arriving settings. The experimental results can stand as an evidence for the reasonableness of the random arrival model. }

\trver{
\begin{table}[t]
\centering
\begin{tabular}{|c|c|c|c|c|}
    \hline
    \multirow{2}{*}{\bf Dataset} & \multicolumn{2}{c|}{\bf Real-World} & \multicolumn{2}{c|}{\bf Random arrival}  \\ \cline{2-5}
    & FIRM & Agenda & FIRM & Agenda \\ \hline
   Digg	& $1.9\times 10 ^{-5}$& $0.008$	& $2.4 \times 10^{-5} $	& $0.015 $ \\ \hline
Flickr	& $1.2 \times 10^{-4}$	& $3.4$	& $1.5 \times 10^{-4}$ 	& $2.4$
\\ \hline
\end{tabular}
\vspace{0mm}
\caption{Update cost under different edge arrival settings.} \label{tab:timegraph_cmp}
\end{table}
}

\header
{\bf Accuracy.} To evaluate the accuracy performance of {\oursolution} and its competitors, we first perform a sufficient number of updates ($5 \mathrm{\sim} 10$ percent edges are inserted) and then measure the relative error of ASSPPR queries. We stop the process if it cannot finish in 48 hours (thus we have no accuracy results for Agenda/{\agendastar} on large graphs). As to FORAsp+, since its accuracy performance is independent of the update process (it always reconstructs the whole index for each update), we can save computational cost by constructing the index only once, after all updates have been applied. \trver{ As for the ground truth of ASSPPR values, we utilize the power method running $160$ rounds and yield a result with a precision at least $0.8^{160} \approx 3.1 \times 10^{-16}$ which is negligible if we store the value in double-precision floating-point type. Recall that the $\epsilon$ relative accuracy guarantee of ASSPPR query holds for those PPR values higher than $\delta = \frac{1}{n}$. For a ASSPPR query, we ignore those nodes with an actual PPR value smaller than $\delta$, and compute the average and maximum relative error among the remaining nodes. }

\begin{figure}[t]
\vspace{-2mm}
    \centering
    \includegraphics[height=2mm]{figures/query_legend.eps}
    \includegraphics[height=2.7cm]{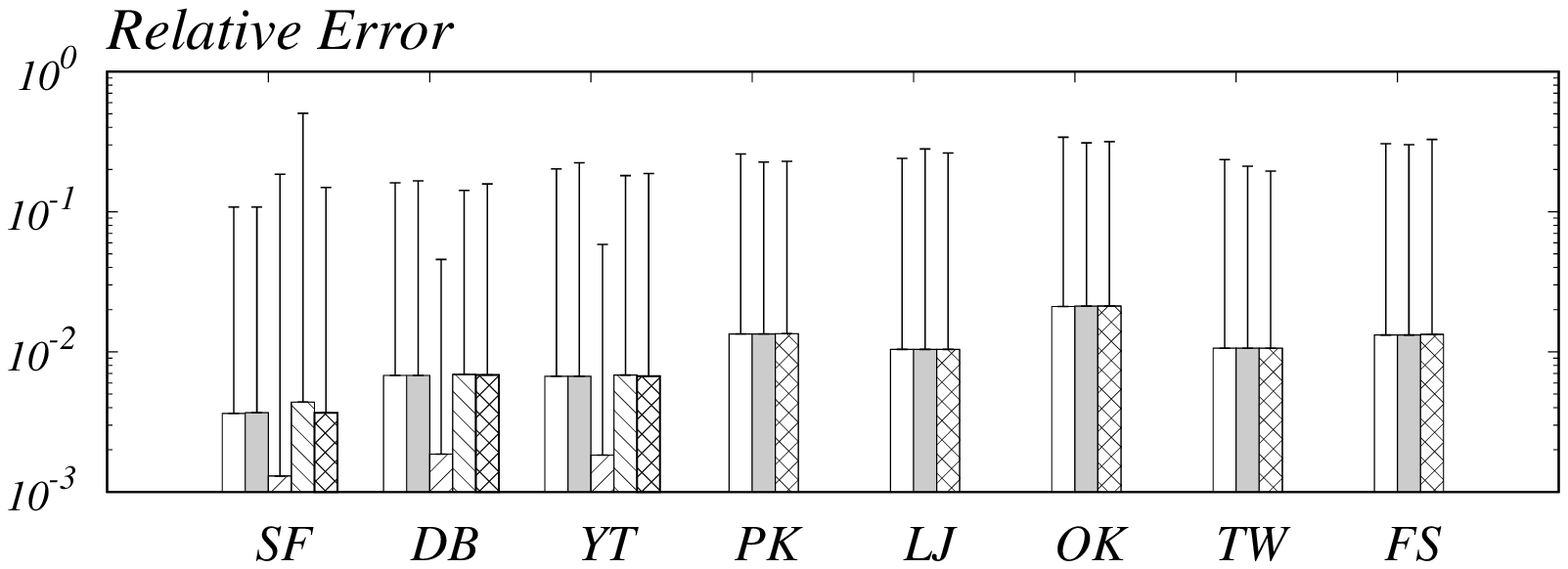}
    \caption{Accuracy results.} \label{fig:perf:precision_hist}
\end{figure}

The experimental results are shown in Figure \ref{fig:perf:precision_hist}, where the box (resp. bar) represents the average (resp. maximum) relative error. {\oursolution} has the identical precision as FORAsp (and FORAsp+), which verifies the correctness of our solution. Besides, as discussed in Section~\ref{sec:sol:dyn}, Agenda has a practical precision higher than other methods which comes from its conservative bound of index inaccuracy and hence a tighter bound in FORA phase. Correspondingly, {\agendastar} gives a precision comparable to (slightly worse than) FORAsp. Even though we make an aggressive assumption on the inaccuracy of its lazy-updated index, which means the precision of {\agendastar} will be theoretically worse than other methods, the gap of average relative error between {\agendastar} and FORAsp is not significant. As we described in Sections~\ref{sec:iprea} and \ref{sec:sol:dyn}, $O(1)$ changes of random walks are adequate to guarantee the precision. However, the lazy-update scheme renews more random walks since it traces the inaccuracy roughly. Therefore, in most cases, the inaccuracy of the lazy-updated index is not as large as Agenda supposed. In summary, our {\oursolution} achieves the best query and update efficiency when providing identical accuracy as alternatives including Agenda, where we tune it as {\agendastar} to gain similar accuracy as {\oursolution}. \crver{In our technical report \cite{firm-tr}, we have also done experiments with only 1000 updates so that Agenda/{\agendastar} can finish in a reasonable time. We have similar observations as discussed above. } \trver{We also run experiments with only 1000 updates so that Agenda and {\agendastar} can finish in a reasonable time. As shown in Figure \ref{fig:perf:precision_1k_hist}, we can see similar observations as discussed above, that is, in terms of accuracy, Agenda is visually better than the other four algorithms (i.e., FORAsp, FORAsp+ FIRM, and {\agendastar}), which have considerably similar average relative error among all datasets. }

\trver{
\begin{figure}[!t]
    \centering
    \includegraphics[height=2.8cm]{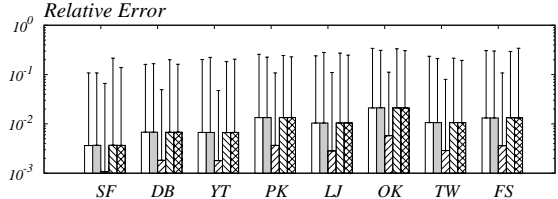}
    \caption{Accuracy results for 1000 updates.} \label{fig:perf:precision_1k_hist}
\end{figure}
}

\begin{figure}[t]
    \centering
    \includegraphics[height=2.7cm]{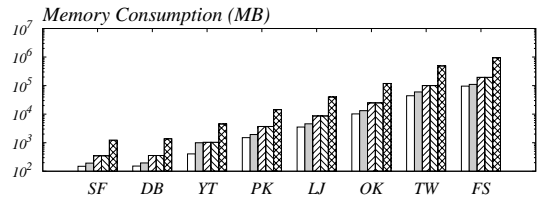}
    \caption{Memory Consumption} \label{fig:perf:mem}
\end{figure}

\header
{\bf Memory Consumption.} Figure~\ref{fig:perf:mem} shows the memory consumption of all methods. FORAsp needs to maintain the evolving graph itself. Then, FORAsp+ costs about 2x memory of FORAsp to store the terminals of pre-sampled random walks, and Agenda/{\agendastar} requires more additional space to store the reverse graph to support Backward-Push. {\oursolution} costs about 8x space as that of FORAsp+ to trace the random walks efficiently. However, with the new sampling scheme introduced in Section~\ref{sec:sample}, {\oursolution} can still handle huge-scale graphs like Twitter and Friendster in memory. In practice, several times more space consumption is an acceptable trade-off for orders of magnitude times speed-up of processing performance because expanding the memory is much easier than enhancing computing power. Moreover, there are several techniques, e.g. $\sqrt{1\mathrm{-}\alpha}$ random walk \cite{WeiHX0SW18}, to make full use of the complete path for more accurate estimation and thus can reduce the space without loss of precision.
\section{Conclusions}
In this paper, we present {\oursolution}, an efficient framework to handle approximate single source PPR problems on evolving graphs. Theoretical analysis proves that our proposal has $O(1)$ time cost for each update in expectation and experiments show that {\oursolution} dramatically outperforms competitors in most scenarios.

\begin{acks}
Sibo Wang is supported by Hong Kong RGC ECS grant (No. 24203419), RGC GRF grant ( No. 14217322), RGC CRF grant (No. C4158-20G), Hong Kong ITC ITF grant (No. MRP/071/20X), NSFC grant (No. U1936205), and a gift from Huawei. Zhewei Wei is supported in part by the major key project of PCL (PCL2021A12), by National Natural Science Foundation of China (No. 61972401, No. 61932001), by Beijing Natural Science Foundation (No. 4222028), by Peng Cheng Laboratory, by Alibaba Group through Alibaba Innovative Research Program, by CCF-Baidu Open Fund (NO. 2021PP15002000) and by Huawei-Renmin University joint program on Information Retrieval.
\end{acks}

\balance
\bibliographystyle{ACM-Reference-Format}
\bibliography{reference}

\appendix

\end{document}